\documentclass[10pt,twocolumn]{article}
\usepackage[textwidth=7in,textheight=8.5in]{geometry}
\usepackage{graphicx}
\usepackage{amsmath} 
\usepackage{amssymb} 
\usepackage{fancyvrb} 
\usepackage{soul}
\usepackage{fancyhdr}
\usepackage{epstopdf}
\usepackage{lineno}
\usepackage{cancel}
\usepackage{subfigure}
\usepackage{hyperref}
\makeatletter
\renewcommand{\maketitle}{\bgroup
\begin{flushleft}
  \begin{Huge}
  \textbf{\@title}\\
  \end{Huge}
  \vspace{1cm}
  \@author
\end{flushleft}\egroup
}
\makeatother
\title{Analysis of Prospective Super-Symmetry Inherent in the $pp$ Collision Data at $7$ TeV from CMS Collaboration Using Novel Two-Dimensional Multifractal-Detrended Fluctuation Analysis Method with Rectangular Scale}
\author{%
    \textbf{{\large Susmita Bhaduri}}$^{1}$, \textbf{{\Large Anirban Bhaduri}}$^{2}$\\
    $^{1}$Moonshot Analytics and Design 3/1/1, Dhakruia Station Lane, Kolkata - 700031, West Bengal, India \\
    $^{2}$Moonshot Analytics and Design 3/1/1, Dhakruia Station Lane, Kolkata - 700031, West Bengal, India\\
    \underline{$^{1}$\textbf{ORCID}:0000-0003-1246-9124, E-mail: susmita.sbhaduri@moonshot.net.in}\\
    \underline{$^{2}$\textbf{ORCID}:0000-0002-7787-3550, E-mail: bhaduri.anirban@moonshot.net.in}\\
}
\begin{document}
\twocolumn[
  \begin{@twocolumnfalse}
    \maketitle
  \end{@twocolumnfalse}
  ]
\noindent

\date{\today}
%
\begin{abstract}
Search for Super-Symmetry in High-Energy-Physics is of enormous interest for the past few decades. Continuous searches were conducted at LHC regarding Super-Symmetry for prompt, non-prompt, R-parity conserving and violating generation and decays. The limits obtained from these analysis to detect the signatures of Super-Symmetric-particles, revealed greater possibilities of such experiments in collider. These signatures are usually derived assuming a bit optimistic conditions of the decaying process of s-particles to final-states. Moreover, Super-Symmetry might have been in a disguised state in lower-mass-scales resulting from challenging mass-spectra and mixed-modes of decays. The proposed chaos-based, novel method of \textit{Two-Dimensional-Multifractal-Detrended-Fluctuation-Analysis((2D)MF-DFA)}~\cite{Yeh2012}, is extended using \textit{rectangular-scale}. The (2D)experimental data-surfaces are constructed using the component-space(in the $X,Y,Z$ co-ordinates) taken out from the $4$-momenta of final-state-signatures of the produced di-muons from the selected events. Two publicly-available datasets are used here. First is the data from MultiJet primary $pp$ collision-data from RunB(2010) at $7$-TeV~\cite{cms2010_multi_7T}, used in analysis of the Super-Symmetry~\cite{Chatrchyan20131,Chatrchyan20132} with Razor-variables. Second is the data from primary-dataset of $pp$ collisions at $7$-TeV from RunA(2011) of CMS-collaboration~\cite{cms2011_7T}. The (2D)Multifractal behaviour of particle-production-process is studied in terms of symmetry-scaling, the inherent scale-freeness and multifractality. The analysis outcome for Super-Symmetric-data is compared with the same for the non-Super-Symmetric-data in terms of the generalized-Hurst-exponent and (2D)Multifractal-spectrum-width. Unusual and significantly different scaling-behaviour and long-range-correlation is observed between the final-state-signatures of the di-lepton production-process of the first and second datasets. This difference may indicate a possible signature of Super-Symmetry which may be missed in the conventional method of analysing the invariant-mass/transverse-momentum-spectrum.
\end{abstract}
%

\textbf{PCAS Nos.:} \textit{10, 11.30.Pb, 24.60.Ky, 24.60.Lz}

\section{Introduction}
\label{intro}
Over a number of past decades, the Standard Model(SM) of particle physics gives us an elaboration of the fundamental particles and their respective interactions~\cite{Glashow1961,Weinberg1967,Glashow1970}. UA1 and UA2 experiments in $1983$ have confirmed that for short-range weak interactions the the $W$ and $Z$ bosons have huge mass but bosons in SM have to be massless to keep SM lagrangian invariant under local gauge transformations. Brout, Englert, Higgs mechanism suggested a probable solution where gauge bosons can gain mass in gauge invariant process through spontaneous symmetry breaking~\cite{Englert1964,Higgs1964,Guralnik1964}. Glashow, Weinberg and Salam have integrated this concept with the SM by formulating the theory of $W$ and $Z$ bosons with huge mass and an additional elementary $0$-spin particle which is the Higgs boson interacting with the gauge bosons and the fermions. The discovery of the Higgs boson by the ATLAS~\cite{Aad2012} and CMS~\cite{Chatrchyan2012} collaborations at the Large Hadron Collider(LHC)~\cite{Evans2008} gave the final approval of the Standard Model. The conjectures of the Standard Model(SM) have been established experimentally with increasing robustness. 

However the Standard Model(SM) has few challenges like gravitational interactions cannot be included in this model, origination of charge parity violation yielding to matter versus antimatter and also the dark matter is not found in this model. Also, it does not answer how neutrinos gain mass etc. There has been search for a framework which unifies all interactions. For resolution of Standard Model(SM)'s limitations many theories have come up.
Among them, Super-Symmetry(SUSY) has been one of the most reasonable extensions of the Standard Model(SM) in the field of particle physics~\cite{Salam1989,Wess1974,Nilles1984,Haber1985,Barbieri1982,Dawson1985,Witten1981,Dimopoulos1981}.
It allows to determine coupling unification at the energy of $10^{16}GeV$. It comes up with a prospective candidate of dark matter - lightest Super-Symmetric particle(LSP)~\cite{Jungman1996}
which is a mandatory element to decode the concept of quantum gravity based on the framework of string theory. 
The signature of Super-Symmetry(SUSY) in the detector, is driven by the nature of the lightest Super-Symmetric particle(LSP) which is dependent on the underlying process of Super-Symmetry(SUSY) breaking.
The Super-Symmetry model eliminates the quadratic divergences in radiative modifications to the Higgs boson mass by default. For each particle in the Standard Model, Super-Symmetry infuses a \textit{super-partner} or the \textit{sparticle}, with a spin difference from the Standard Model particle by $1/2$ unit. There exist theoretical propositions which suggest that the masses of \textit{sparticles} might be less than $\sim 1TeV$~\cite{Witten1981,Dimopoulos1981}
and hence the experiments at the Large Hadron Collider(LHC) are best suited to be analyzed for the discovery of Super-Symmetry particles. 
For the past decades considerable amount of missing transverse energy has been considered as the most rigorous observable to identify the production and decay of Super-Symmetry(SUSY) particles at colliders and quantity of missing transverse energy depends on the division of mass among the heavier sparticles.
Till date, usual di-lepton SUSY searches in CMS~\cite{Chatrchyan20121,Chatrchyan20133} needed several jets with large transverse momentum($p_T$) correlated with large values of $H_T$, the scalar sum of the transverse momenta of all the jets satisfying the jet selection criteria, and absence of large transverse energy to distinguish a SUSY signal from the very large SM backgrounds. In the work~\cite{Chatrchyan2013}, comparatively moderate criteria of the absence of transverse energy($\cancel{E}_t>40GeV$) and also $H_T(H_T>120GeV)$, has been used.
The Standard Model(SM) has been extended with softly broken Super-Symmetry(SUSY)~\cite{Golfand1971iw,Ramond1971,Volkov2007,Fayet1975,Jungman1996}
to predict new elementary particles which are \textit{super-partner} of the Standard Model(SM) particles. The Fermilab Tevatron~\cite{Abazov2008,Aaltonen2009}[7,8]
and the CERN LHC~[eg.~\cite{Aad2012c,Aad2012a,Aad2012b,Chatrchyan20123}]
have concentrated on the events having signatures of high-energy hadronic jets and also of leptons decayed from \textit{squarks} and \textit{gluinos} produced in pairs, considering the assumption of \textit{R-parity}~\cite{Farrar1978} conservation. 
These events usually have high degree of missing transverse energy resulting from the weakly interacting \textit{super-partners} which are stable and one of the \textit{super-partners} is generated in each of the two decay chains. 
In these final states, searches for Super-Symmetry(SUSY) were conducted by the ATLAS~\cite{Aad2014} at $13$TeV and CMS~\cite{Khachatryan2014} Collaborations at $8$TeV.
Chatrchyan et al.~\cite{Chatrchyan20131,Chatrchyan20132} have applied the razor approach to search for Super-Symmetric particles, after considering the \textit{R-parity}~\cite{Farrar1978} conservation.
In this work they have considered two analyses - a comprehensive search for new heavy-particle-pairs which decay to final states with at least two jets and missing transverse energy and an exclusive search for the final states having at least one jet arising from a bottom quark. For both the analyses the final states of hadronic, single-lepton and di-lepton events from the MultiJet primary dataset of the RunB(2010) of $pp$ collision at $7$TeV from CMS collaboration~\cite{cms2010_multi_7T}, are considered. 
Chatrchyan et al.~\cite{Chatrchyan20131,Chatrchyan20132} have analyzed chosen events in the two-dimensional razor plane of $M_R$ - razor kinematic variable, defined with regards to the momentum of the two megajets and $R$, a variable without dimensions signifying the missing transverse energy. These razor variables are established on the generic procedure of the production of two heavy particles in pair, each one decaying to an unidentified particle with visible decay products. The two-dimensional shape analysis in razor-variable plane is validated with simulated events, and no remarkable excess over the background expectations has been noticed. The output of the search for Super-Symmetry~\cite{Chatrchyan20131,Chatrchyan20132} has been used to drive simplified Super-Symmetry models~\cite{Alwall2009,Alwall2009a,Alves2012}.

A significant feature of multiparticle production process is the inherent fluctuation. Correlation analysis may yield valuable information about the dynamics of this process. In the past, for quite a few years, fluctuation and correlation had been analyzed extensively using novel perspective of studying non-statistical fluctuation which had resulted in rigorous interpretation of the pionisation process. 
To begin with, Bialas and Peschanski~\cite{bialash986} introduced a novel concept to analyze multiplicity fluctuations with respect to scaled factorial moments to identify and analyze the pattern of large non-statistical fluctuation, and finally heading towards the physical explanation of the origin of the fluctuations. They suggested that the character of the factorial moments is similar to the process of intermittency observed in the hydrodynamics of turbulent fluid flow. Intermittency is a process which manifests prominent local fluctuations in consistent and large statistical systems.
Moreover, it was noticed that multipion production process manifests a power-law trend of the factorial moments in regards to the magnitude of phase-space intervals in decreasing mode~\cite{bialash986}.
Bialas and Peschanski~\cite{bialash986} also suggested a connection between intermittency and fractal behaviour. Initially fractal pattern of multi-pion production process was analyzed from the context of intermittent fluctuations by utilizing the technique of factorial moment. Then a simple relationship was found between the intermittency indices and anomalous fractal dimension~\cite{bia1ash1988,dewolf1996}. The inherent cascading process in the multipion production process naturally generates a fractal structure. Also the scale invariance existing in the process of hadronization was obvious from the spectrum of fractal dimensions.  
After that, innumerable methods based on the fractals had been applied to study the multipion production process using the parameters of Gq moment and Tq moment~\cite{hwa90,paladin1987,Grass1984,hal1986,taka1994}. Then different approaches like the Detrended Fluctuation Analysis(DFA), multifractal-DFA(MF-DFA) method~\cite{Peng1994,kant2002} were implemented exhaustively for studying non-stationary, nonlinear data series to detect their long-range correlations.
In the recent years, various studies based on the multi-fractal properties of the process of particle production had been reported~\cite{Albajar1992,Suleymanov2003,YXZhang2007,Ferreiro2012}. In a number of recent works, self-similarity had been probed in the field of particle physics, like - in the Jet and Top-quark generation process at the experiments of Tevatron and LHC~\cite{Tokarev2015}, in the process of strangeness production in $pp$ collisions at the experiments of RHIC~\cite{Tokarev2016}, in the proton spin phenomena and asymmetry of jet production process~\cite{Tokarev20151} to explain the collective phenomena~\cite{Baldina2017} and in the implementation of the concept of self-similar symmetry to dark energy~\cite{TomohideSonoda2018}.

A fundamental change in approach occurred with the latest advancements in the area of research using complex-network based methodologies. Albert and Barab{\'{a}} made remarkable contributions in this area applying the analytic models and tools for small-world, random and also scale-free graphs~\cite{Albert2002,Barabasi2011}. 
Lacasa et al. developed the Visibility Graph methodology~\cite{laca2008,laca2009} which attained prominence because of its completely novel and accurate perspective to estimate the fractal behaviours. 
Studies had been conducted in the area of the multiplicity fluctuation process in nucleus-nucleus and hadron-nucleus interactions, applying this Visibility Graph methodology in few recent works~\cite{Bhaduri20167,Bhaduri20171,Bhaduri20165,Bhaduri20166,Bhaduri20172,Bhaduri20181,Bhaduri20183,Bhaduri20182,Bhaduri20184,Bhaduri20191}.
Also, Pb-Pb VSD masterclass data at $2.76 TeV$ per nucleon pair from ALICE Collaboration~\cite{alice} was used for scaling analysis of the pseudorapidity space, using both the method of complex network based Visibility Graph and multifractal-DFA(MF-DFA)~\cite{Peng1994,kant2002}, to study the prospective phase transition and the signature of QGP in some latest works~\cite{Bhaduri2630203,Bhaduri20184}. 
In a latest work~\cite{Bhaduri20201}, the fluctuation pattern inherent in the dynamics of particle production process in high energy collision has been analyzed using the multifractal-DFA(MF-DFA) method and also multifractal-detrended-cross-correlation(MF-DXA) analysis using the pseudorapidity values of di-muon data taken out from the $pp$ collision at $7$TeV and $8$TeV respectively from CMS collaboration~\cite{cms2011_7T,cms2012_8T}.
A recent review work had been reported about the complexities involved in resonance production for various high energy collisions like $pp$, $pA$ and $AA$ collisions at LHC (using data from ALICE collaboration), to understand the complexity and origin of different resonance states and eventually understand the inherent dynamics of the particle production process and the properties of the produced particles for the various collision systems~\cite{Werner2018}.

Chatrchyan et al.~\cite{Chatrchyan2013} have presented a search for Super-Symmetry by applying an ANN (Artificial Neural Network) model to differentiate prospective Super-Symmetry signals from the background of Standard Model. They implemented the search for Super-Symmetry in the events having pairs of oppositely signed leptons in the final state, hadronic jets and missing transverse energy, from the data of Run-A(2011) of the $pp$ collision at $7$TeV at CMS detector. They have confirmed accordance between the expectations and observations of the Standard Model which gives rise to more optimized Super-Symmetric Standard models. Such successful attempts in this area of analysis of pionisation process in high energy interaction using chaos-based fractal, multi-fractal and complex-network procedure including attempts to probe the origin of the resonance states ~\cite{Werner2018} as well as the use of new computational techniques such as neural networks in the area of high energy collisions have encouraged and inspired us to extend the methods to analyze another significant area of high energy interaction using a novel multi-fractal method.

In this work we have attempted to analyze the inherent symmetry scaling in the di-lepton production process using a novel method of extended two-dimensional(2D) MF-DFA method using rectangular scales. The resultant dataset (in totality) analyzed for the search of Super-Symmetry using razor filter~\cite{Chatrchyan20131,Chatrchyan20132} generated from the MultiJet primary $pp$ collision data from RunB of 2010 at $7$TeV from the CMS collaboration~\cite{cms2010_multi_7T} has been used for this analysis. The symmetry scaling for di-muon production process in case of Super-Symmetry is compared with the same for the non-Super-Symmetric di-muon data produced in $pp$ collision in RunA of 2011 at $7$ TeV from CMS collaboration~\cite{cms2011_7T}, using the same method of (2D) MF-DFA. The scaling analysis is based on the component-space(in the $X,Y,Z$ co-ordinates) taken out from the $4$-momenta of final state signatures of the produced di-muons, obtained from the data from CMS collaboration~\cite{cms2010_multi_7T,cms2011_7T}. The signature of Super-Symmetry may be identified from the remarkable differences in the scaling behaviour of the di-muon production process between the two kinds of data from CMS collaboration, indicated by the Hurst exponent, generalized Hurst exponent and the width of (2D) MF-DFA spectrum, without using conventional invariant mass or transverse momentum techniques.

The rest of the paper is structured as follows. Section~\ref{ana} describes the methods of analysis in detail and the importance of various properties of scale-freeness. Inside the Section~\ref{exp}, the Section~\ref{data} elaborates the details of the data and Section~\ref{method} describes the detailed steps of our analysis and the relevant inferences. The test results are further analyzed with regards to the the physical relevance of the proposed parameters with respect to the conventional ones for analysing Suer-Symmetry in the high energy interactions in Section~\ref{con}.

\section{Method of analysis}
\label{ana}
Fractals are geometric pattern which is reiterated over various scales to generate self-similar patterns~\cite{mandelbrot1983fractal}. Natural artifacts like coastal lines, profile of trees, leaves, clouds or a mountain etc. are proved to be fractals. Magnifying a small part of a fern leave, generates similar geometrical pattern like the whole fern leave, which shows that they are self-similar. Most significant property of fractals is the reiteration of the geometric pattern over different scales, which is known as the self-similarity. Fractal dimension is a ratio signifying a non-statistical index of complexity, comparing detailed change of a fractal pattern along with the scale. 
Fractals are categorized into two types: mono and multi-fractals. For mono-fractal systems, the scaling properties are consistent across different scales but for multi-fractals these properties are different for different regions and are more complex in nature. Multi-fractal objects have a range of different non-integer dimensions. Various methods that have been suggested to measure Fractal Dimension - Wavelet Transform Modulus Maxima (WTMM), Fluctuation Analysis (FA), Detrended Fluctuation Analysis (DFA), Detrended Moving Average (DMA), Multi-fractal Detrended Fluctuation Analysis (MF-DFA) and Hurst exponent, Power of Scale-freeness of a Visibility Graph (PSVG) etc.

In the present paper we have extended two-dimensional (2D) Multifractal-Detrended Fluctuation Analysis (MF-DFA) proposed Yeh et al.~\cite{Yeh2012}, by selecting rectangular scale for doing Multifractal analysis for experimental data surfaces. To elaborate the methodology we have briefly explained the Detrended Fluctuation Analysis (DFA) and Multifractal Detrended Fluctuation Analysis (MF-DFA) method for single dimensional data series first~\cite{Peng1994,Kantelhardt2001,kant2002} and then elaborated the extended (2D) (MF-DFA) method in detail in the Section~\ref{2dmfdfa}. 

\subsection{Two-dimensional (2D) MF-DFA algorithm}
\label{2dmfdfa}

The DFA algorithm implemented on an one dimensional data series, measures self-similar or fractal-like correlations by calculating the scaling pattern of the Root-Mean-Square fluctuation of both the integrated and detrended time series~\cite{Peng1994,Kantelhardt2001,kant2002}. Let us denote time series with $N$ samples as $x(i)$ for $i = 1,2, \ldots N$ and average of the series is denoted by $\bar{x} = \frac{1}{N}\sum_{i=1}^{N} x(i)$. Then the integrated random walk series $y(i)$ is calculated as per the equation~\ref{eq:l-dfa}.
\begin{eqnarray}
\label{eq:l-dfa}
y(i) = \sum_{k=1}^{i} [x(k)-\bar{x}], i = 1,2,\ldots,N  
\end{eqnarray}

Then, series $y(i)$ is detrended for a particular scale of $s$ by dividing $y(i)$ with $N_s$ number of non-overlapping segments of length $s(N_s \equiv int(N/s))$ and then subtracting the local trend of the each segment derived from least-square fitting for that segment, denoted by $x_v(i)$ for segment sequence $v$ corresponding to a specific value of $s$. For each $s$, the detrending function $F^2(s,v)$ of the integrated series $y(i)$ for a particular segment $v = 1,2, \ldots N_s$ is calculated as per the equation~\ref{eq:l-dfa1}.

\begin{equation}
\label{eq:l-dfa1}
    F^2(s,v) = \frac{1}{s}\sum_{i=1}^{s} \{y[(v-1)s+i]-x_v(i)\}^2
\end{equation}

with different $s$ and corresponding $v \in 1,2,\ldots,N_s$.

Then, the $q^{th}$-order function of fluctuation, $F_q(s)$, is calculated by averaging the values of $F^2(s,v)$ computed for all the set of segments-$v = 1,2, \ldots N_s$ produced for each $s$, as per the equation~\ref{eq:l-dfa2}.
\begin{eqnarray}
\label{eq:l-dfa2}
	F_q(s) = \left\{\frac{1}{N_s}\sum_{v=1}^{N_s} [F^2(s,v)]^{\frac{q}{2}}\right\}^{\frac{1}{q}}
\end{eqnarray}

For $q = 2$, computation of $F_2(s)$ for different values of $s$ would correspond to conventional method of Detrended Fluctuation Analysis(DFA)~\cite{Peng1994,Kantelhardt2001,kant2002}. If the data series $x(i)$ is long range power correlated, then $F_q(s)$ vs $s$ for a specific $q$, will display power-law behaviour and $\log_{2} [F_q(s)]$ would depend on $\log_{2} s$ in a linear fashion, where the slope for $q=2$ corresponds to the so-called \textbf{Hurst exponent}~\cite{Peng1994,Kantelhardt2001,kant2002} and the slope for other non-zero values of $q$ is denoted as the generalized Hurst exponent. After extending Detrended Fluctuation Analysis (DFA) for monofractal data series, Multifractal Detrended Fluctuation Analysis (MF-DFA) method has been proposed for multifractal data series and the width of the multifractal spectrum denotes the magnitude of the multifractality for self similar data series~\cite{Peng1994,Kantelhardt2001,kant2002}.

DFA and MF-DFA~\cite{Peng1994,Kantelhardt2001,kant2002} methods were proposed for one-dimensional self-similar data series but in this experiment scaling analysis is based on the components of the momemtum($GeV$) in the $X,Y$ and $Z$ co-ordinate of the produced di-leptons. So, for this analysis, a two dimensional (2D) MF-DFA algorithm is applied~\cite{Yeh2012}, where the data is represented as a two dimensional matrix and the scale is chosen as square matrix segment. However, in most real-life and natural data surfaces, their two-dimensional matrix representations are not perfect square. Hence here we propose a (2D) MF-DFA analysis for self-similar surfaces using a rectangular scale.

The proposed method is elaborated in the below steps.
\begin{enumerate}
\item \label{random_y}Consider $X$ is a self-similar surface of dimension $M \times N$ where $X(i,j)$ denotes the value of $i^{th}$ row and $j^{th}$ column. So, like equation.~\ref{eq:l-dfa}, the integrated random walk surface or the profile surface $Y$ is calculated as:

\begin{equation}
    Y(i,j) = \sum_{m=1}^{i}[ X(m,j) - {\bar{X}}_{jc}]  + \sum_{m=1}^{j}[ X(i,m) - {\bar{X}}_{ir}]
\end{equation}

for $1 \leq i \leq M$, $1 \leq j \leq N$. Where, $Y(i,j)$ is the value $(i,j)$-th element on the integrated matrix, $X(m,j)$ is the $m^{th}$ element on the $j^{th}$ column, $X(i,m)$ is the $m^{th}$ element on the $i^{th}$ row. ${\bar{X}}_{jc}$ is the average of the values in $j^{th}$ column and ${\bar{X}}_{ir}$ is the average of the values of $i^{th}$ row.

\item \label{partition_in_scales} The integrated surface represented in matrix $Y(i,j)$ is partitioned into $M_{ms} \times N_{ns}$ disjoint rectangular surfaces of the size $ms \times ns$. Here, $M_{ms} = [\frac{M}{ms}]$, $N_{ns} = [\frac{N}{ns}]$ and $ms_{min} \leq ms \leq ms_{max}$, $ns_{min} \leq ns \leq ns_{max}$. $ms \times ns$ can be regarded as the scale, denoted by $S$, for which the fluctuation is measured.

\item\label{scale_chosing} There are a number of factors to choose the maximum and minimum size of the scale $S$ i.e. $ms\times ns$ as we need to repeat the below steps varying the values of $S$ and this is important for the computation of local fluctuations for each segments. Different arguments, statistical and phenomenological exist for choosing the maximum and minimum segment size~\cite{kant2002}. The statistical argument is to choose minimum and maximum segment sizes that provide a numerical stable estimation of variance function and different orders of fluctuation function for each surface segment. Further, it is desirable to have a equal spacing between scales when they are arranged in a log-log scale i.e. $log(S)$ versus the $q^{th}$ order fluctuation function for an optimal performance of the linear regression to estimate the $q^{th}$ order Hurst exponent $H_q$. 
Considering the above we choose the below
\begin{itemize}
\item $ms_{max}$ as mentioned in step~\ref{partition_in_scales} as $\frac{M}{4}$ is optimal to provide sufficient number of segments in the computation.
\item $ns_{max}$ as mentioned in step~\ref{partition_in_scales} as $\frac{N}{4}$ is optimal to provide sufficient number of segments in the computation.
\item $ms_{min}$ as mentioned in step~\ref{partition_in_scales} as $16$ is optimal to prevent over-fitting.
\item $ns_{min}$ as mentioned in step~\ref{partition_in_scales} as $16$ is optimal to prevent over-fitting.
\item Scale resolutions $S_{res}$ i.e. the number of different scale sizes for which we repeat the above steps is chosen as $S_{res} = 19$.
\item To have equal spacing between scale sizes following is done
\begin{itemize}
\item We calculate $\hat{exp_{ms}} = linspace\{log_2(ms_{min}), \\ log_2(ms_{max}),S_{res}\}$
where $\hat{exp_{ms}}$ is a vector with $S_{res}$ elements placed at equal interval in log scale between $log_2(ms_{min})$ and $log_2(ms_{max})$.
\item Similarly $\hat{exp_{ns}} = linspace\{log_2(ns_{min}),\\ log_2(ns_{max}),S_{res}\}$ where $\hat{exp_{ns}}$ is a vector with $S_{res}$ elements placed at equal interval in log scale between $log_2(ns_{min})$ and $log_2(ns_{max})$.
\item We thereby get a vector $\hat{ms}$ of size $S_{res}$ according to the equation $\hat{ms} = \lfloor {2^{\hat{exp_{ms}}}}\rfloor$ for each element in $\hat{exp_{ms}}$.
\item Similarly a vector $\hat{ns}$ of size $S_{res}$ is obtained as per the equation $\hat{ns} = \lfloor {2^{\hat{exp_{ns}}}}\rfloor$ for each element in $\hat{exp_{ns}}$ 
\item Thus a new vector of scales $\hat{S}$ is calculated from $\hat{ms}$ and $\hat{ns}$ where $S(i) = ms(i) \times ns(i)$ and $i \in [1,2 \ldots S_{res}]$.
\end{itemize}
\end{itemize}

\item \label{division_in_scales} For a particular $S(i) \in \hat{S}$ and corresponding $ms(i) \in \hat{ms}$, $ns(i) \in \hat{ns}$, the full integrated surface $Y(i,j)$ can be partitioned into smaller surface segments denoted by $Y_{k,l}$ of size $ms(i) \times ns(i)$ such that each element in surface $Y_{k,l}$ can be denoted by $Y_{k,l}(o,p)$ mapped to the original surface represented by $Y(i,j)$ such that  ---
\begin{itemize}

\item $0 \leq k < M_{ms}$ and $0 \leq l < N_{ns}$, where $M_{ms}$ and $N_{ns}$ are calculated as per the particular $ms(i)$ and $ns(i)$.
\item $1+(k \times ms(i)) \leq o \leq (k + 1) \times ms(i)$, $1+(l \times ns(i)) \leq p \leq (l + 1) \times ns(i)$ i.e. here the generic sequence numbers denoted by $o$ and $p$, starts from $1+(k \times ms(i))$ and $1+(l \times ns(i))$ respectively for particular values of $k$ and $l$ and are incremented by $1$ until $o$ and $p$ reach the values of $(k + 1) \times ms(i)$ and $(l + 1) \times ns(i)$ respectively.
Eg. for $k=0$ and $l=1$, $o = 1, 2, \dots ms(i)$ and $p = 1+ns(i), (1+ns(i))+1, \dots (2 \times ns(i))$. This way set of $Y_{k,l}$ for different values of $k$ and $l$ make up the original surface $Y(i,j)$ for the particular scale $S$.

\item Each surface segment $Y_{k,l}$ in the full surface $Y(i,j)$, thereby becomes a separate matrix of size $ms(i) \times ns(i)$ for each scale $S \in \hat{S}$ and corresponding $ms(i) \in \hat{ms}$ and $ns(i) \in \hat{ns}$. Also each element of the independent surface segment denoted by $Y_{kI,lI}$, can be denoted by $Y_{kI,lI}(x,y)$ with $x = 1,2 \dots ms(i)$ and $y = 1,2 \dots ns(i)$, for $1 \leq kI \leq M_{ms}$ and $1 \leq lI \leq N_{ns}$, where $M_{ms}$ and $N_{ns}$ are calculated as per the particular $ms(i)$ and $ns(i)$ corresponding a specific scale $S(i) \in \hat{S}$ as defined in the step~\ref{scale_chosing}. 
\end{itemize}

\item\label{regr_calc}For each individual surface segment $Y_{kI,lI}$ as defined in the step~\ref{division_in_scales}, corresponding to each scale $S(i) \in \hat{S}$, $ms(i) \in \hat{ms}$, $ns(i) \in \hat{ns}$ and $i \in [1,2 \ldots S_{res}]$ as defined in the step~\ref{scale_chosing}, we can obtain a local trend $\tilde{Y}_{kI,lI}$ by fitting it with a pre-chosen bi-variate polynomial function. The simplest function could be a plane in equation~\ref{eq:bivariate_plane}. In this work, we have adopted this trending function.
\begin{equation}
\label{eq:bivariate_plane}
    \tilde{Y}_{kI,lI}(x,y) = Ax+By+C 
\end{equation}
where $x = 1,2 \dots ms(i)$ and $y = 1,2 \dots ns(i)$ and $A,B$ and $C$ are parameters to be determined and can be estimated easily through simple matrix operations, derived from the least-square method.
 
\item\label{resi_calc} For each $Y_{kI,lI}$ the residual matrix $R_{kI,lI}$ obtained by subtracting the least-square fitted bi-variate polynomial function for the particular surface segment, can be determined as per equation \ref{eq:residual_matrix}, for a particular scale $S(i) \in \hat{S}$.
\begin{equation}
\label{eq:residual_matrix}
R_{kI,lI}(x,y) = [Y_{kI,lI}(x,y) - \tilde{Y}_{kI,lI}(x,y)]
\end{equation}

\item\label{rms_calc} For each scale $S(i) \in \hat{S}$, and corresponding $ms(i) \in \hat{ms}$, $ns(i) \in \hat{ns}$ for $i \in [1,2 \ldots S_{res}]$, de-trending is performed as per equation \ref{eq:detrended_fluctuation_function}, by subtracting the least-square fitted surface segment $\tilde{Y}_{kI,lI}(x,y)$ from the corresponding actual surface segment $Y_{kI,lI}(x,y)$ to deduce the variance.
\begin{equation}
\label{eq:detrended_fluctuation_function}
F^2_{kI,lI}(S(i)) = \frac{1}{s_n}\sum_{x=1}^{ms(i)}\sum_{y=1}^{ns(i)}{[R_{kI,lI}(x,y)}]^2
\end{equation}
where count of elements in $Y_{kI,lI}(x,y)$ is denoted by $s_n = ms(i) * ns(i)$.
This variance $F^2_{kI,lI}(S(i))$ is calculated for all the surface segments $Y_{kI,lI}$, where $kI=1,2, \ldots M_{ms}$ and $lI=1,2, \ldots M_{ns}$ and $M_{ms}$ and $N_{ns}$ are calculated for each of the $ms(i)$ and $ns(i)$ corresponding to a particular $S(i) \in \hat{S}$. Hence for a particular $S(i)$, there would be $M_{ns} * N_{ns}$ number of $F^2_{kI,lI}(S(i))$-s for the same number of surface segments $Y_{kI,lI}$ for the full integrated surface $Y(i,j)$.
\\

The steps~\ref{regr_calc},~\ref{resi_calc},~\ref{rms_calc} are repeated each $S(i) \in \hat{S}$ where $i=1,2 \ldots S_{res}$ and for each scale a set of $M_{ns} * N_{ns}$ number of $F^2_{kI,lI}(S(i))$-values, is calculated.

\item \label{qfluc} Then the $q^{th}$ order fluctuation of fluctuation, denoted by $F_q(S(i))$ is calculated by averaging the values of $[F^2_{kI,lI}(S(i))]^\frac{q}{2}$ corresponding to $F^2_{kI,lI}(S(i))$-values calculated for all surface segments $Y_{kI,lI}$ as per step~\ref{rms_calc}, where $i=1,2 \ldots S_{res}$. This $q^{th}$ order fluctuation of fluctuation function is defined as per the equation \ref{eq:qth_order_fluctuation}. 
\begin{equation}
\label{eq:qth_order_fluctuation}
F_q(S(i)) = \left[ \frac{1}{M_{ms}*N_{ns}}\sum_{kI=1}^{M_{ms}}\sum_{lI=1}^{N_{ns}}[F^2_{kI,lI}(S(i))]^\frac{q}{2} \right]^{ \frac{1}{q}}
\end{equation}
where 
\begin{itemize}
\item $M_{ms}$ and $N_{ns}$ are as calculated for each of the $ms(i) \in \hat{ms}$ and $ns(i) \in \hat{ns}$ corresponding to each scale $S(i) \in \hat{S}$ with $i=1,2 \ldots S_{res}$, as per the method specified in step~\ref{partition_in_scales} and~\ref{scale_chosing}.
\item For $q=0$ the values of $q^{th}$ order fluctuation of fluctuation - $F_q(S(i))$ are not affected by the values of segments with small and large values of variance function $F^2_{kI,lI}(S(i))$. Also for $q=0$ the $\frac{1}{q}$ would blow up to infinity. Hence, for $q=0$, $F_0(S(i))$ is calculated as per the equation\ref{eq:fluctuation_for_0}
\begin{equation}
\label{eq:fluctuation_for_0}
F_0(S(i)) = e^{\left[0.5*\frac{1}{M_{ms}*N_{ns}}* \log(F^2_{kI,lI}(S(i)))\right]}
\end{equation}
for each scale $S(i) \in \hat{S}$.
\item For $q = 2$, computation of $F_2(S(i))$ for all the values of $S(i) \in \hat{S}$ would correspond to conventional method of Detrended Fluctuation Analysis (DFA)~\cite{Peng1994}.
\item In this experiment $q$ varies from $(-5)$ to $(+5)$.
\end{itemize}

\item \label{seqn4}After executing the above steps, it is observed that for a particular $q$, $F_q(S(i))$ increases with increasing $S(i)$. It must be noted here that each scale $S(i)$ is actually the area of corresponding rectangular scale with dimensions of $ms(i)$ and $ns(i)$, which amounts to $ms(i) * ns(i)$. If the data surface is long range power correlated, then $F_q(S(i))$ vs $S(i)$ for a specific $q$, will display power-law behaviour as per the equation[~\ref{eq:scale_vs_fluctuation}].
\begin{eqnarray}
\label{eq:scale_vs_fluctuation}
F_q(S(i)) \propto S(i)^{h(q)}
\end{eqnarray}
where $S(i) \in \hat{S}$ with $i=1,2 \ldots S_{res}$.
If this type of scaling exists then $\log_{2} [F_q(S(i))]$ would depend on $\log_{2} S(i)$ in a linear fashion, where $h(q)$ is the slope which is dependent on $q$. Here $h(2)$ is analogous to the so-called \textbf{Hurst exponent}~\cite{Kantelhardt2001} and $h(q)$ is termed as the generalized Hurst exponent.

\item \label{seqn5}The scaling trend of the variance function $F^2_{kI,lI}(S(i))$ is similar for all surface segments in case of a monofractal surface. To put differently, the averaging of $F^2_{kI,lI}(S(i))$ would exhibit consistent scaling pattern for various values of $q$ and hence for monofractal surfaces $h(q)$ becomes independent of $q$. 

But, if small and large fluctuations in the surface have differing scaling pattern, then $h(q)$ becomes significantly dependent on $q$. In these scenarios, for values of $q>0$, $h(q)$ depicts the scaling behaviour of the surface segments with large fluctuations and for values of $q<0$, $h(q)$ depicts scaling pattern of the surface segments with smaller fluctuations. 
The generalized Hurst exponent $h(q)$ for a multifractal surfaces is related with the classical multifractal scaling exponent $\tau(q)$ as per the equation~\ref{eq:multifractal_scaling_exp}.
\begin{equation}\label{eq:multifractal_scaling_exp}
\tau(q) = qh(q)-1 
\end{equation}
\item \label{seqn6} As multifractal surface has a range of Hurst exponents, $\tau(q)$ depends upon $q$~\cite{ASHKENAZY200319} in a nonlinear fashion. The singularity spectrum, here denoted by$f(\alpha)$, is related to $h(q)$ as per the equation[~\ref{eqn6}].
\begin{eqnarray}
\alpha = h(q)+qh'(q), f(\alpha) = q[\alpha-h(q)]+1 
\label{eqn6}
\end{eqnarray}

Here the singularity strength is represented by $\alpha$ and $f(\alpha)$ represents the dimension of the subset of surface segments corresponding to $\alpha$. Different values of $f(\alpha)$ corresponding to different values of $\alpha$ result into multifractal spectrum of $f(\alpha)$ which forms an arc and the difference between the maximum and minimum values of $\alpha$ for this spectrum, is the \textbf{width of the multifractal spectrum} or the magnitude of the multifractality of the input self-similar surface, here denoted by $X$.

\item For $q=2$, if $h(q)$ or $h(2)=0.5$ there is no correlation in the data surface. $h(2) > 0.5$ infers that a high value of $X(i,j) \in X$ is followed by another large value in the surface or there is existence of persistent long-range cross-correlations in the surface. However if $h(2) < 0.5$, there must be anti-persistent long-range correlations existing in the surface suggesting a high value of $X(i,j) \in X$ would possibly be followed by a small value in the surface and vice verse.

\end{enumerate}

\section{Experimental details}
\label{exp}
Two experimental primary datasets which are made publicly available by CMS collaboration are used for the the proposed analysis. The detailed specification of the data is given in Section~\ref{data} and the detailed method of the experiment is elaborated step by step in Section~\ref{method}.
\subsection{Data description}
\label{data}

From the CMS collaboration, the MultiJet primary dataset of $pp$ collisions in AOD format from RunB of 2010 at $7$ TeV~\cite{cms2010_multi_7T}, is extracted and  considered as the source data-set of first level, for this experiment. During data selection procedure all the runs recorded by CMS collaboration, are qualified as \textit{good} for physics analysis if all sub-detectors, trigger (an algorithm to select particle collisions to be stored in the primary dataset and rest are discarded), lumi and physics objects like electron, muon, photon, jet etc. show the expected performance. First level of validation is done based on the evaluations of off-line shifters and then the feedback is taken from the detector and experts from Physics Object Group(POG). All these information is stored in a specific database named Run Registry. The CMS Data Quality Monitoring group validate the integrity of the certification and creates a (.json) file of qualified runs to be utilized for physics analysis. The (.json) file \href{https://doi.org/10.7483/OPENDATA.CMS.YDT4.BW6J}{\textit{\underline{link1}}}, describe luminosity sections containing the \textit{good} runs which have been selected from the primary dataset of events for further processing. This is the second level of selection. 
Then the filter is run over the MultiJet primary dataset from the CMS open data, after the second level of selection, and the events with jet candidates with certain parameters having the required threshold values are selected and resultant files in (.root) and (.csv) formats are created. The jets are reconstructed and grouped into two mega-jets - one leading mega-jet with largest transverse momentum($p_T$) and another subleading mega-jet with largest transverse momentum($p_T$). The mega-jets are derived as sum of the $4$-momenta of their constituent particles. From the reconstructed jets from the events with large degree of missing transverse energy, the razor variables $M_R$ and $R^2$ are calculated to be used in Super-Symmetric particle searches~\cite{Chatrchyan20132}. This is the final level of selection. In this analysis we have used this set of $X,Y$ and $Z$ components taken out from the $4$-momenta of final state signatures of the constituent di-muons from the pair of mega-jets. This is the data space for Super-Symmetric data. 

Similarly, the primary dataset of $pp$ collisions at $7$ TeV from RunA of 2011~\cite{cms2011_7T} of the CMS collaboration, in AOD format is taken as another source dataset in this experiment. Events stored in this dataset were selected due of the presence of precisely two \textit{global} muons in the event with specific range of invariant mass. This is the first level of selection. Similar to the case of MultiJet primary dataset from CMS collaboration, here also the (.json) file is created after validating the certification by the CMS Data Quality Monitoring group, is gives in the link - \href{https://doi.org/10.7483/opendata.cms.3q75.7835}{\textit{\underline{link2}}} for $7$ TeV. This (.json) file describes the luminosity sections containing the \textit{good} runs which have been selected from the primary dataset for further processing. This is the second level of selection. Here also, the output containing di-muon event information is extracted in both (.root) and (.csv) format from the collision datasets.
Then the component-space(in the $X,Y,Z$ co-ordinates) taken out from the $4$-momenta of final state signatures of the produced di-muons, is extracted from output dataset generated from the $pp$ collisions data at $7$ TeV from CMS collaboration~\cite{cms2011_7T}. This is the data space for non-Super-Symmetric data.

In this analysis we have used these two set of component-spaces(in the $X,Y,Z$ co-ordinates) taken out from the $4$-momenta of final state signatures of the produced multi particle data - one Super-Symmetric and another non-Super-Symmetric data.

%

\subsection{Data analysis and results}
\label{method}
\subsubsection{Data Preparation}
\label{matrices}

The two sets of component-space(in the $X,Y,Z$ co-ordinates) taken out from the $4$-momenta of final state signatures of the produced multi particle data, are extracted from the datasets created out of the MultiJet primary dataset from RunB of 2010 at $7$ TeV~\cite{cms2010_multi_7T} for supersymmetry data and the primary dataset of $pp$ collisions at $7$ TeV from RunA of 2011~\cite{cms2011_7T} for $pp$ collision data from the CMS collaboration for non-Super-Symmetric data. 

For each of the $2$ data surfaces(one Super-Symmetric and another non-Super-Symmetric) two-dimensional matrix is constructed as per the definition of $X$, elaborated in the step~\ref{random_y} in the Section~\ref{2dmfdfa}. Here, each element of the matrix defined as $X(i,j)$, in the step~\ref{random_y} in the Section~\ref{2dmfdfa}, where $i$ is the value of the $X$-component of the $4$-momenta of final-state-signature of the \textit{first} muon of the produced di-muons of a single extracted event, $j$ is the value of the $Y$-component of the $4$-momenta of final-state-signature of the same muon and the value of the element $X(i,j)$ is the value of the $Z$-component of the $4$-momenta of final-state-signature of the same muon. For each element of the matrix $X(i,j) \in X$, $i$ denotes the sequence of the row and $j$ denotes the sequence of column and $X(i,j)$ is the value of the element. The element next to $X(i,j)$ in the matrix would be defined by the $X,Y$ and $Z$ components of the $4$-momenta of final-state-signature of the \textit{second} muon of the produced di-muons of the same extracted event. 

This way, the matrix $X$ would be defined with the $X,Y$ and $Z$ components of the $4$-momenta of final-state-signature of the produced di-muons from each extracted event for each of the $2$ datasets. Each of the $2$ matrices corresponding to $2$ datasets, is of dimension $M \times N$ where 
\begin{itemize}
\item $M=($Maximum of the values of $X$-component of the $4$-momenta of final-state-signature of the produced di-muons for a dataset$)-($Minimum of the values of $X$-component$)+1$.
\item $N=($Maximum of the values of $Y$-component of the $4$-momenta of final-state-signature of the produced di-muons for a dataset$)-($Minimum of the values of $Y$-component$)+1$.
\end{itemize}

For each of the $2$ data surfaces (one Super-Symmetric and another non-Super-Symmetric) of $X,Y$ and $Z$ components of the $4$-momenta of final-state-signature of the produced di-muon data extracted from the MultiJet primary dataset from RunB of 2010 at $7$ TeV~\cite{cms2010_multi_7T} and the primary dataset of $pp$ collisions at $7$ TeV from RunA of 2011~\cite{cms2011_7T} for $pp$ collision data from the CMS collaboration, $3$ sets of $2$ two-dimensional matrices are defined following the above instructions, as described below.

\begin{enumerate}
\item Two-dimensional matrix is constructed as per the definition of $X$, defined in the step~\ref{random_y} the Section~\ref{2dmfdfa}, where for each element $X(i,j)$, $i=$ the value of the $X$-component, $j=$ the value of the $Y$-component and $X(i,j)=$ value of the corresponding $Z$-component. These are denoted by $[x,y,z]_{ss}$ and $[x,y,z]_{pp}$.
\item Two-dimensional matrix is constructed as per the definition of $X$, defined in the step~\ref{random_y} the Section~\ref{2dmfdfa}, where for each element $X(i,j)$, $i=$ the value of the $X$-component, $j=$ the value of the $Z$-component and $X(i,j)=$ value of the corresponding $Y$-component. These are denoted by $[x,z,y]_{ss}$ and $[x,z,y]_{pp}$.
\item Two-dimensional matrix is constructed as per the definition of $X$, defined in the step~\ref{random_y} the Section~\ref{2dmfdfa}, where for each element $X(i,j)$, $i=$ the value of the $Y$-component, $j=$ the value of the $Z$-component and $X(i,j)=$ value of the corresponding $X$-component. These are denoted by $[y,z,x]_{ss}$ and $[y,z,x]_{pp}$.
\end{enumerate}

Once the all the two-dimensional, rectangular matrices are constructed for each of the experimental dataset, the (2D) MF-DFA analysis is done and the width of (2D) MF-DFA spectrum is deduced as per the method elaborated in Section~\ref{2dmfdfa}, for each of the matrix data. The findings and inferences are listed as follows.
\subsubsection{Observations and inferences}
\begin{enumerate}
\item The $q^{th}$ order fluctuation of fluctuation $F_q(S(i))$ of each of the surfaces (here represented by matrices) is calculated as per the equation~\ref{eq:qth_order_fluctuation} of the step~\ref{qfluc} of the (2D) MF-DFA methodology as elaborated in Section~\ref{2dmfdfa} corresponding to each scale $S(i) \in \hat{S}$ with $i=1,2 \ldots S_{res}$. The Figure~\ref{fqs_xyz}-(a) and~\ref{fqs_xyz}-(b) show the trend of $log_2[F_q(S(i))]$ vs $log_2[S(i)]s$ for $q = -5,0,5$, computed for $[x,y,z]_{ss}$ and $[x,y,z]_{pp}$ for all the scales $S(i) \in \hat{S}$ with $i=1,2 \ldots S_{res}$. Similarly, the Figure~\ref{fqs_xzy}-(a) and~\ref{fqs_xzy}-(b) show the same for $[x,z,y]_{ss}$ and $[x,z,y]_{pp}$ and the Figure~\ref{fqs_yzx}-(a) and~\ref{fqs_yzx}-(b) show the same for $[y,z,x]_{ss}$ and $[y,z,x]_{pp}$.

Their linear trend of $log_2[F_q(S(i))]$ vs $log_2[S(i)]s$ for all the values of $q$ establishes the power law trend of $F_q(S(i))$ versus $S(i)$ for the $q^{th}$ orders. This trend in turn confirms the self-similarity and long range power correlation of different orders of the experimental data surfaces. This is true for both Super-Symmetric and non-Super-Symmetric $pp$ collision data at $7$ TeV, which is evident from the Figures.
\begin{figure*}[h]
\centerline{
\includegraphics[width=3.5in]{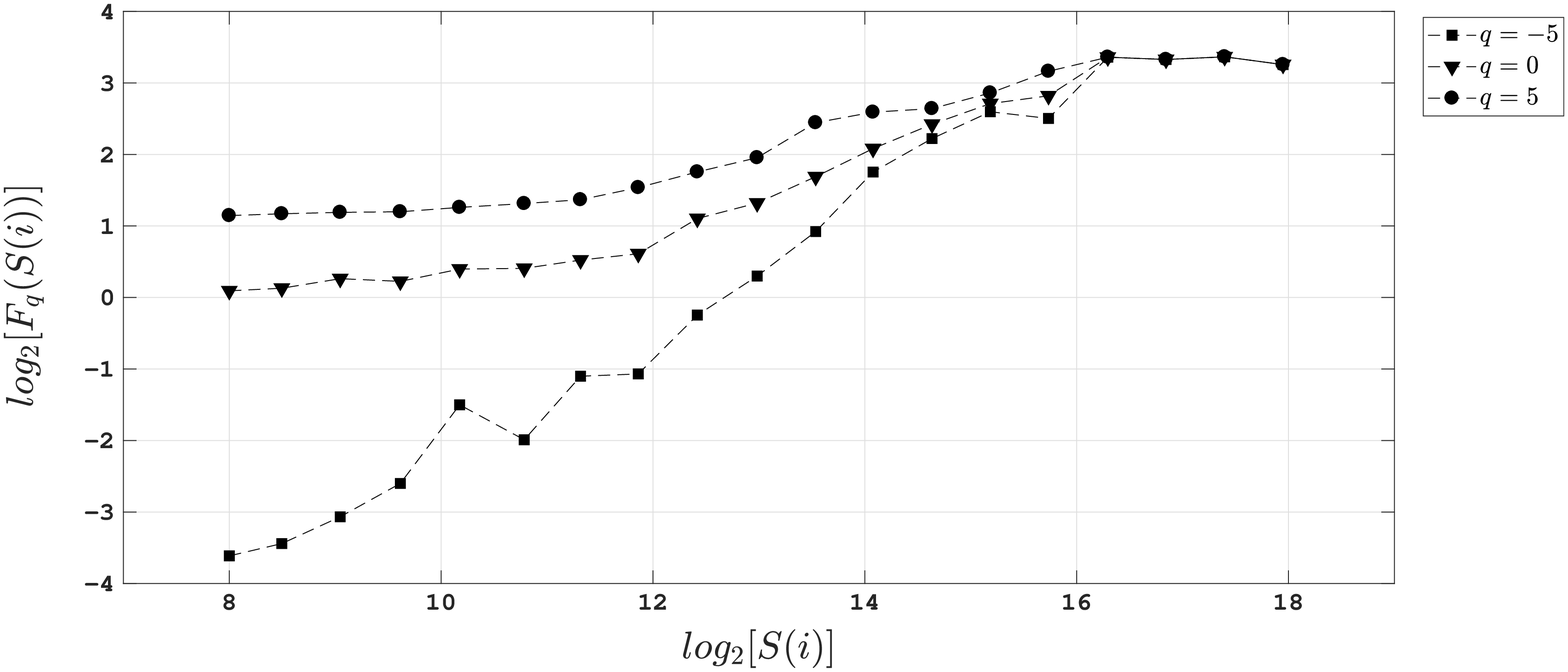}
\includegraphics[width=3.5in]{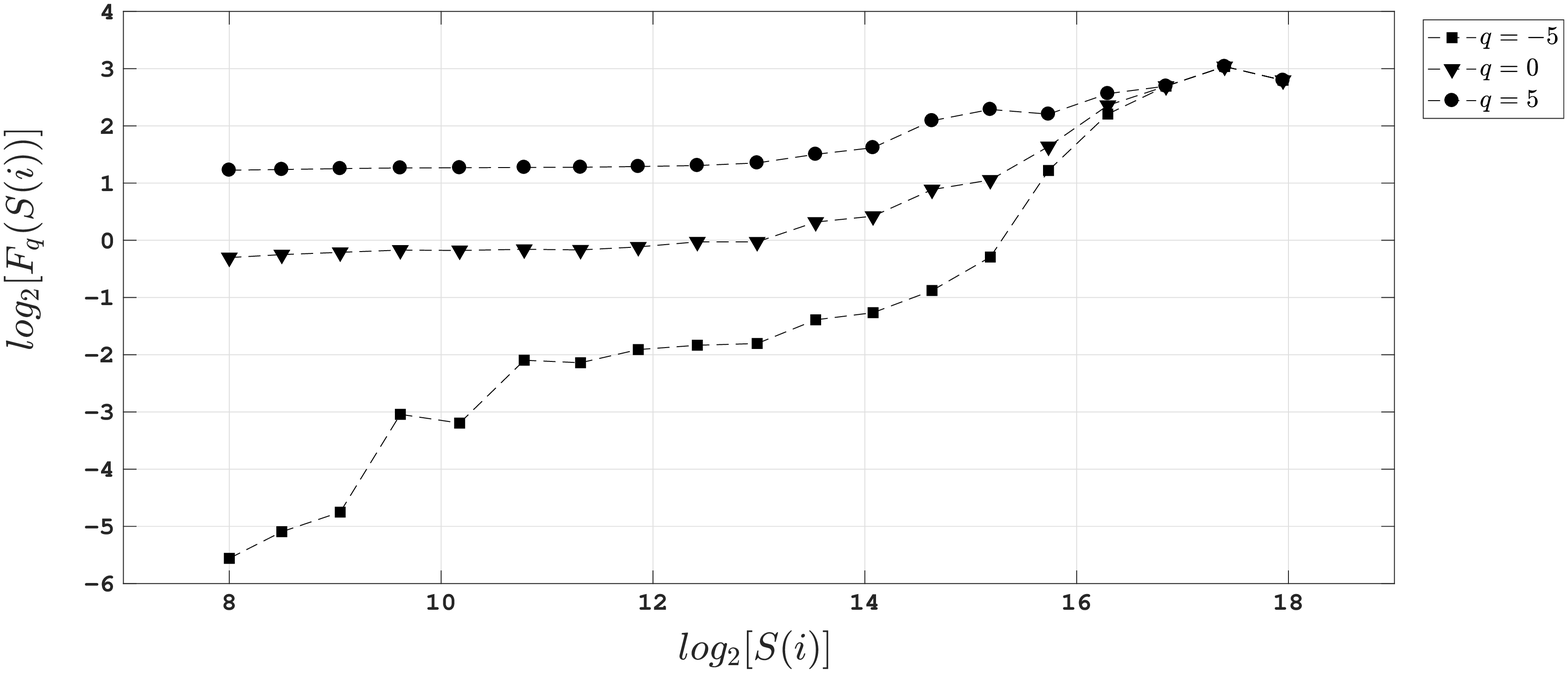}
}
\centerline{(a) \hspace*{6cm} (b)}
\caption{Trend of $log_2[F_q(S(i))]$ vs $log_2[S(i)]s$ for $q = -5,0,5$, computed for (a) $[x,y,z]_{ss}$ (b) $[x,y,z]_{pp}$.}
\label{fqs_xyz}

\centerline{
\includegraphics[width=3.5in]{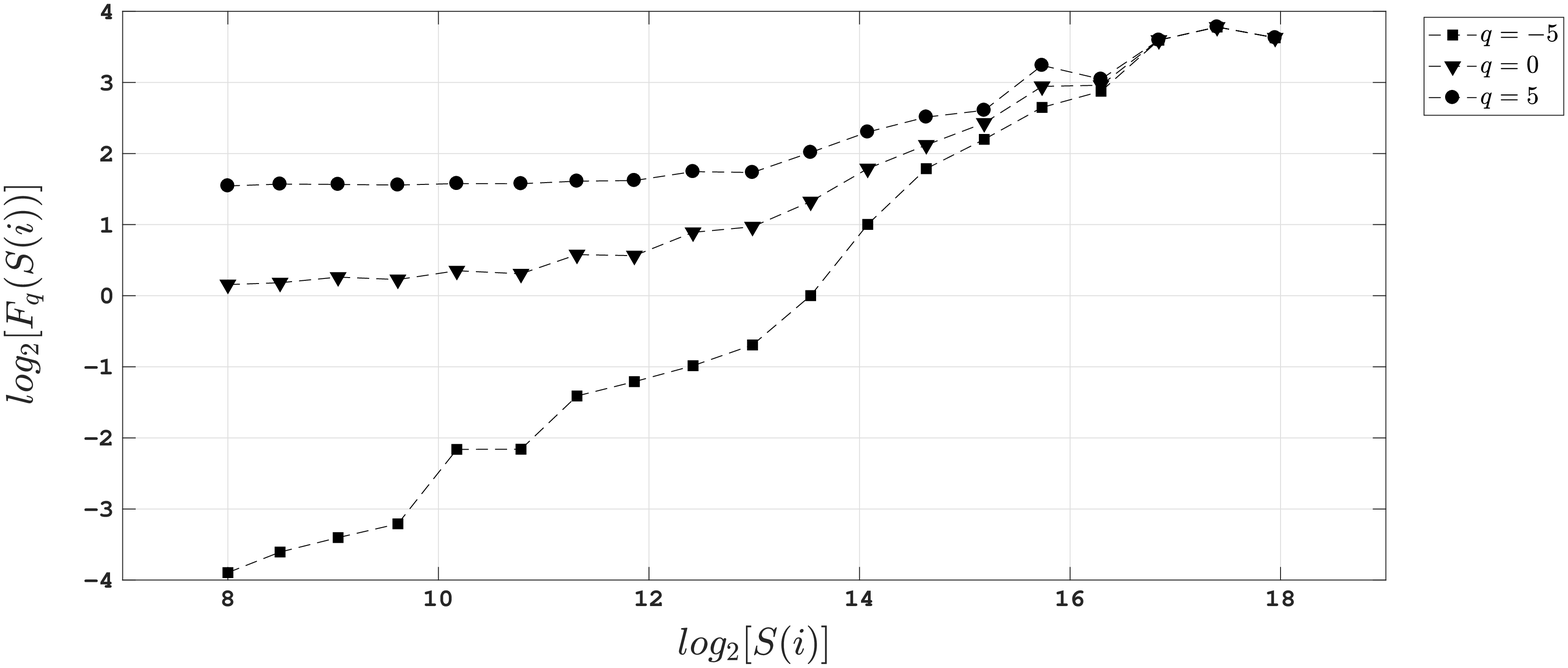}
\includegraphics[width=3.5in]{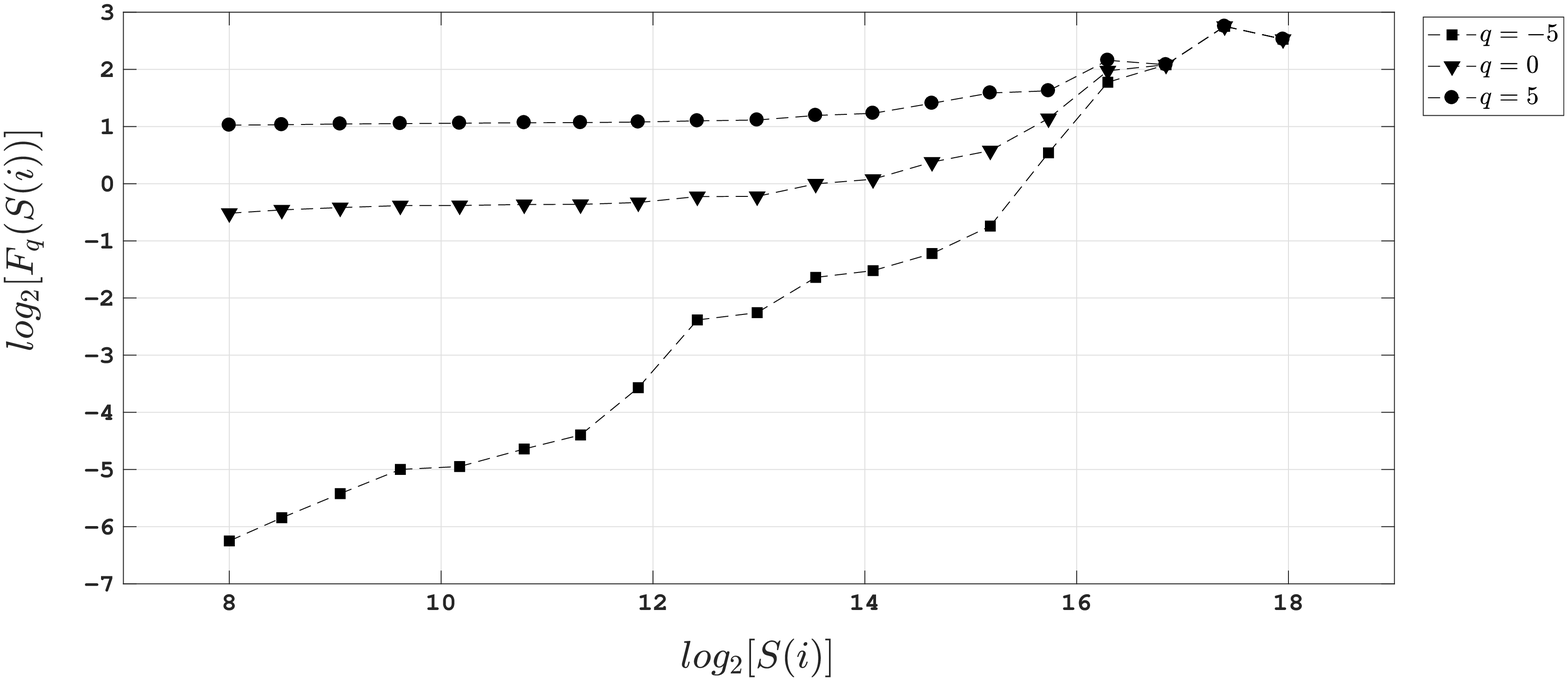}
}
\centerline{(a) \hspace*{6cm} (b)}
\caption{Trend of $log_2[F_q(S(i))]$ vs $log_2[S(i)]s$ for $q = -5,0,5$, computed for (a) $[x,z,y]_{ss}$ (b) $[x,z,y]_{pp}$.}
\label{fqs_xzy}

\centerline{
\includegraphics[width=3.5in]{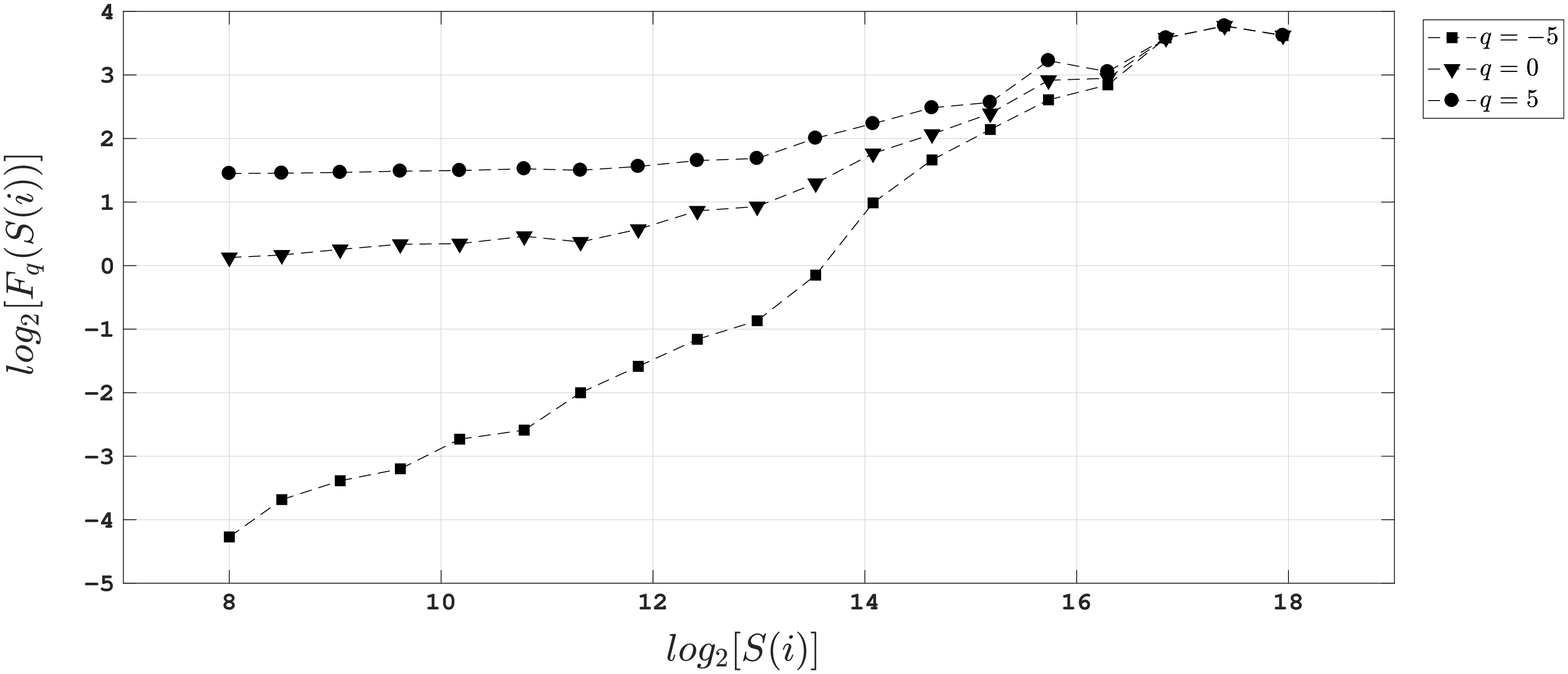}
\includegraphics[width=3.5in]{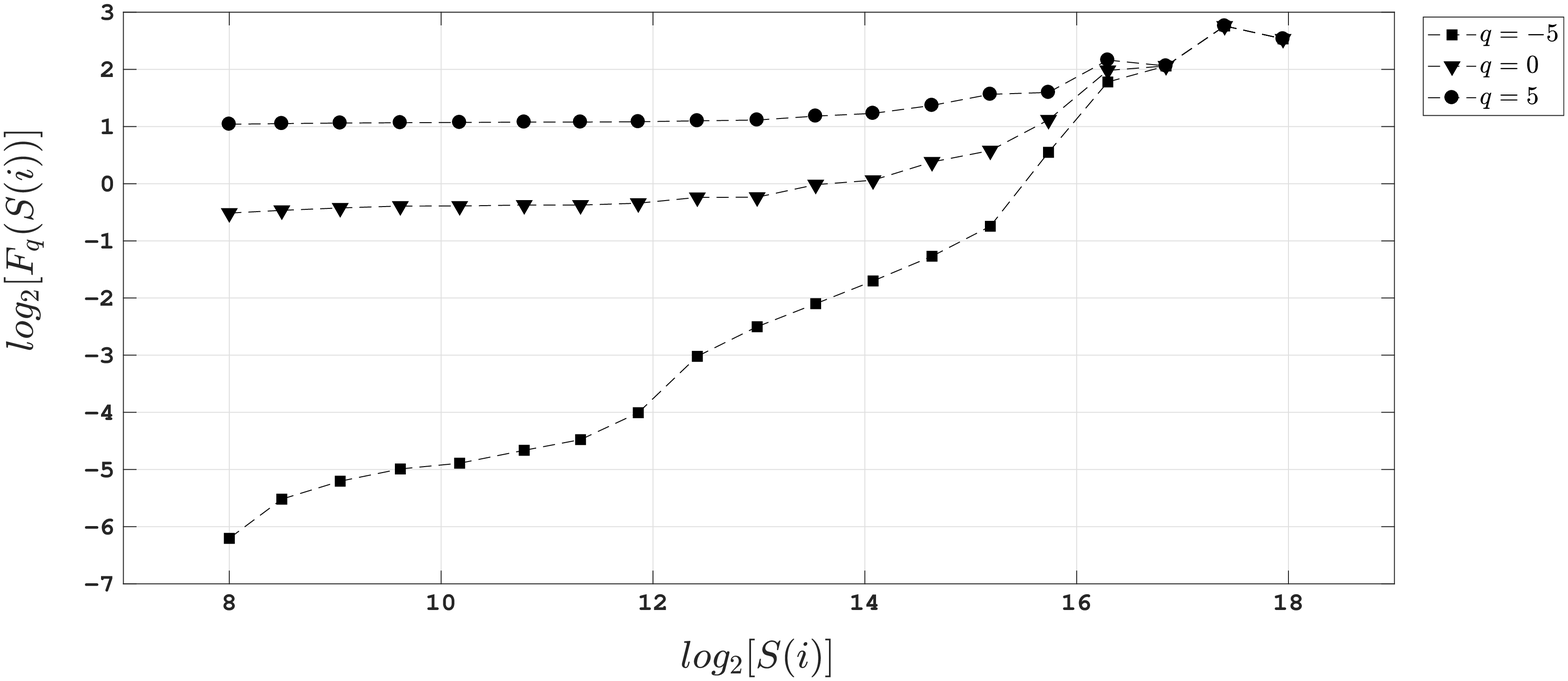}
}
\centerline{(a) \hspace*{6cm} (b)}
\caption{Trend of $log_2[F_q(S(i))]$ vs $log_2[S(i)]s$ for $q = -5,0,5$, computed for (a) $[y,z,x]_{ss}$ (b) $[y,z,x]_{pp}$.}
\label{fqs_yzx}

\end{figure*}

\item As already elaborated in step~\ref{seqn4} of the (2D) MF-DFA methodology in Section~\ref{2dmfdfa}, that for $q = 2$, computation of $F_2(S(i))$ for all the values of $S(i) \in \hat{S}$ would correspond to conventional method of Detrended Fluctuation Analysis(DFA)~\cite{Peng1994} and if the trend of $F_q(S(i))$ versus $S(i)$ is in accordance with the power law, then the slope of $\log_{2} [F_2(S(i))]$ versus $\log_{2} S(i)$ with straight-line fitting, denoted by $h(2)$ is the \textbf{Hurst exponent}~\cite{Kantelhardt2001} and the slope of $\log_{2} [F_q(S(i))]$ versus $\log_{2} S(i)$ with straight-line fitting, denoted by $h(q)$ is the generalized Hurst exponent.

\begin{table*}[h]
\refstepcounter{table}
\label{hurst_comp}
\begin{center}
Table \arabic{table}
Hurst exponent calculated for $6$ experimental data surfaces(matrices)[$3$ for Super-Symmetric data and $3$ for non-Super-Symmetric $pp$ collision data at $7$ TeV] and their randomized versions.
\begin{tabular}{@{}ccc@{}} 
\hline
\multicolumn{3}{c}{\textbf{Hurst Exponent}}\\
\hline
&\textbf{$[x,y,z]_{ss}$}&\textbf{$[x,y,z]_{pp}$}\\
\cline{2-3}
\textbf{Original}&$0.320 \pm 0.020$&$0.160 \pm 0.020$\\
\hline
\textbf{Randomized}&$0.011 \pm 0.002$&$0.009 \pm 0.001$\\
\hline	
&\textbf{$[x,z,y]_{ss}$}&\textbf{$[x,z,y]_{pp}$}\\
\cline{2-3}
\textbf{Original}&$0.250 \pm 0.020$&$0.130 \pm 0.020$\\
\hline
\textbf{Randomized}&$0.013 \pm 0.002$&$0.008 \pm 0.001$\\
\hline
&\textbf{$[y,z,x]_{ss}$}&\textbf{$[y,z,x]_{pp}$}\\
\cline{2-3}
\textbf{Original}&$0.250 \pm 0.020$&$0.130 \pm 0.020$\\
\hline
\textbf{Randomized}&$0.013 \pm 0.002$&$0.008 \pm 0.001$\\
\hline
\end{tabular} 
\end{center}
\end{table*}

For each of the matrices corresponding to $6$ data surfaces ($3$ Super-Symmetric and $3$ non-Super-Symmetric $pp$ collision data at $7$ TeV) constructed as per the method described in step~\ref{matrices} in the current section, a randomized or shuffled version of data surface or matrix is generated and for all of them. The randomization is done by keeping the $Z$ values the same and randomizing the matrix co-ordinates they are placed into. Then the same (2D) MF-DFA analysis is done and the width of (2D) MF-DFA spectrum is calculated. Hurst exponent calculated for each of the data surfaces and their randomized version is listed in the Table~\ref{hurst_comp}. 

From the Table~\ref{hurst_comp}, it's evident that for the data surfaces constructed out of $X$ from $3$ perspectives  - $[x,y,z]$, $[x,z,y]$, $[y,z,x]$, as mentioned in the step~\ref{matrices}, that Hurst exponents or $h(2)$-s are $<0.5$. This signifies that for both data surfaces(Super-Symmetric and non-Super-Symmetric) contain anti-persistent long-range correlations. This means, high value of $X(i,j) \in X$ would possibly be followed by a small value in the surface and vice verse for all the data surfaces constructed for the $3$ perspectives.

Further all Hurst exponents calculated for experimental surfaces are substantially different from those calculated for the randomized ones. This signifies that the inherent anti-persistent long-range correlation and also self-similarity of the data surfaces indicated by the power law behaviour of the data surfaces with respect to the scale, in the experimental surface is not the consequence of shuffling or randomization. Rather these are consequence of the inherent dynamics of the data surfaces. Hence this noticeably different scaling behaviour between experimental and randomized data establishes the statistical significance of the result of the experiment.

Lastly, it's also noted in the Table~\ref{hurst_comp}, that the values of the Hurst exponent calculated for the Super-Symmetric surfaces is significantly more(around $50\%$) than those of non-Super-Symmetric surfaces. This indicates that the anti-persistent long-range correlation and self-similarity inherent in the data surfaces changes significantly from non-Super-Symmetric surfaces to the Super-Symmetric ones and this is an unusual observation arising out of the dynamics of Super-Symmetric data.

\item For each of the $3$-pairs data surfaces (one Super-Symmetric and another non-Super-Symmetric for each pair) and their corresponding randomized versions, the widths of the (2D) multifractal spectrum are calculated as per the steps~\ref{seqn4},~\ref{seqn5} and~\ref{seqn6} following the (2D) MF-DFA method described in Section~\ref{2dmfdfa}. The values of the the widths of the (2D) multifractal spectrum of the experimental and their randomized version for each pair of the data surfaces are compared. 

\begin{figure*}[h]
\centerline{
\includegraphics[width=3.5in]{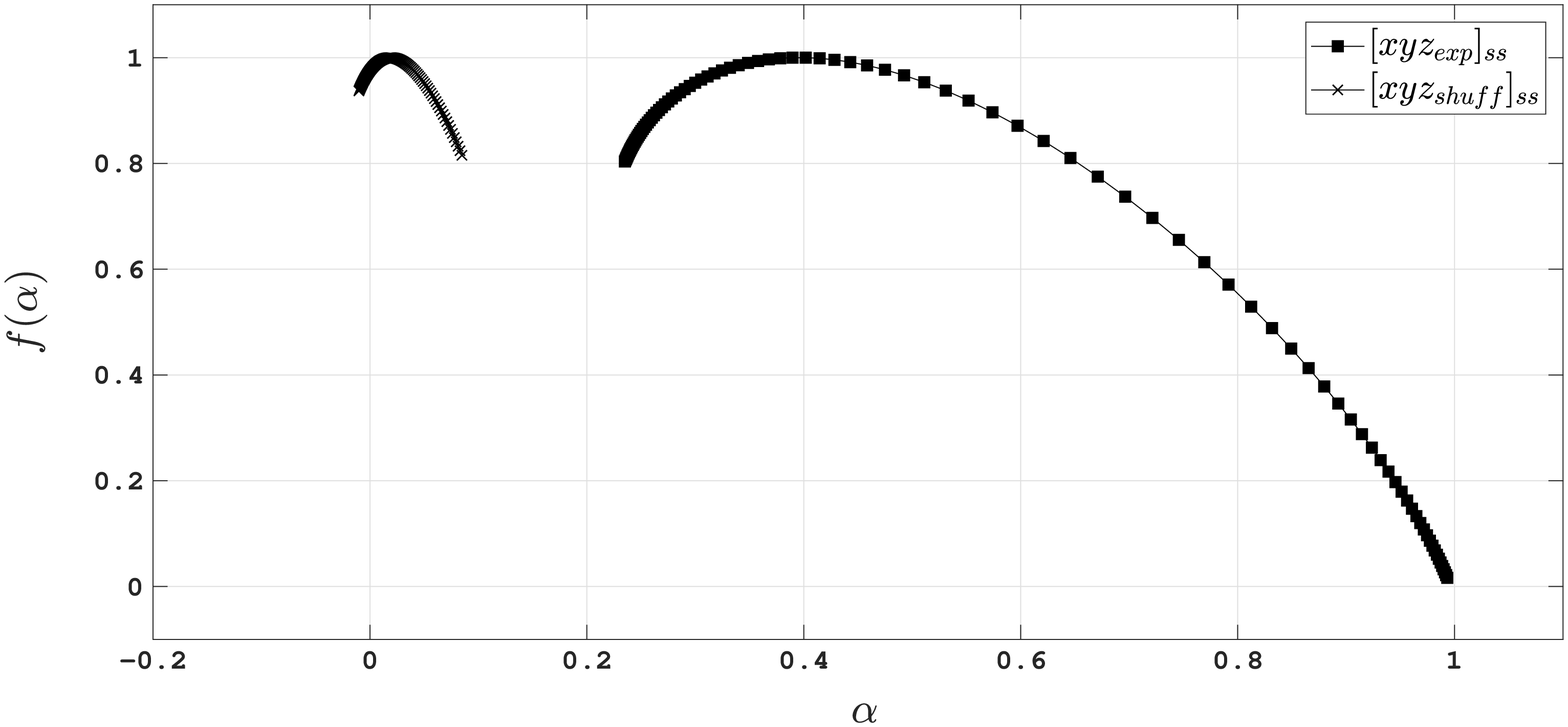}
\includegraphics[width=3.5in]{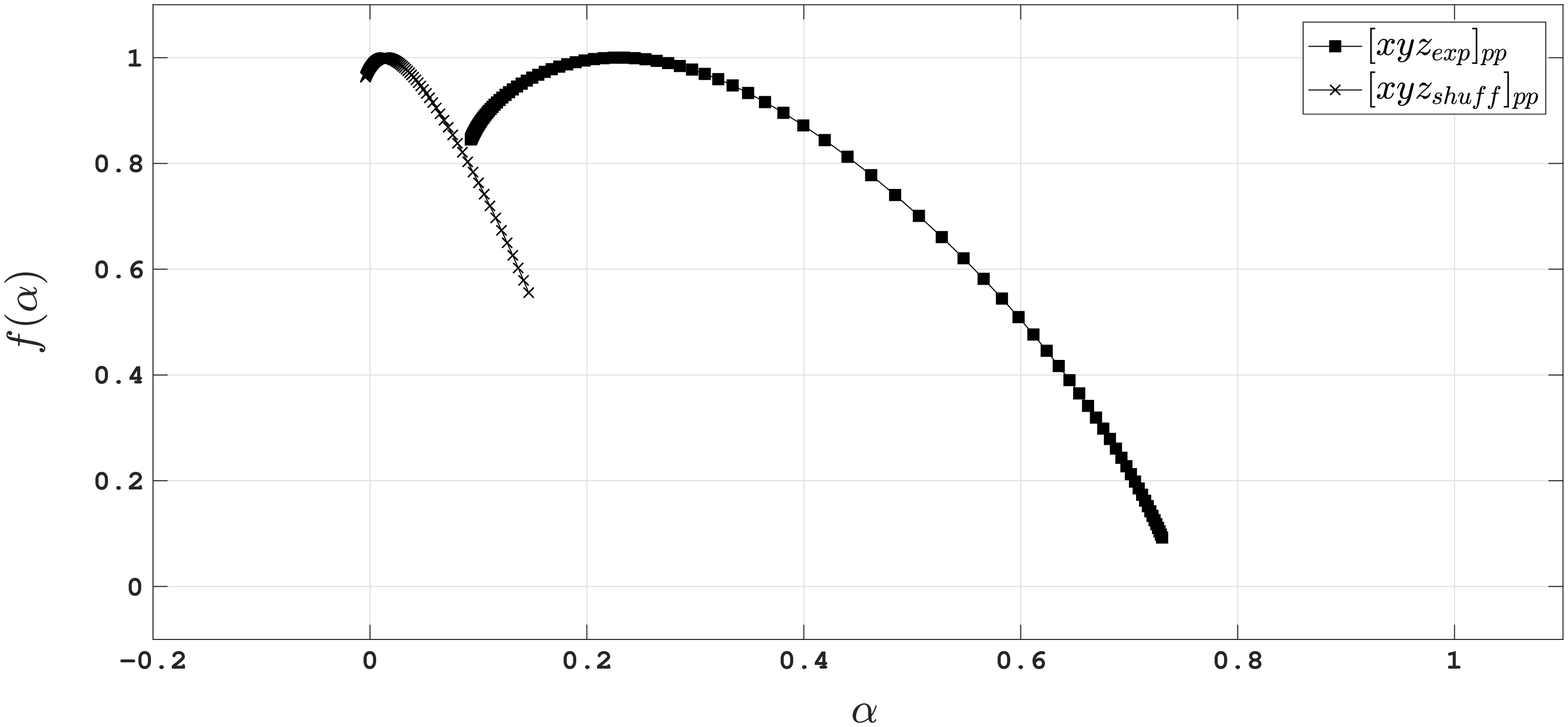}
}
\centerline{(a) \hspace*{6cm} (b)}
\caption{Comparison of the trend of different values of $f(\alpha)$ versus $\alpha$ between the experimental and the randomized version for the data surfaces (a) $[x,y,z]_{ss}$ (b) $[x,y,z]_{pp}$.}
\label{w_xyz}

\centerline{
\includegraphics[width=3.5in]{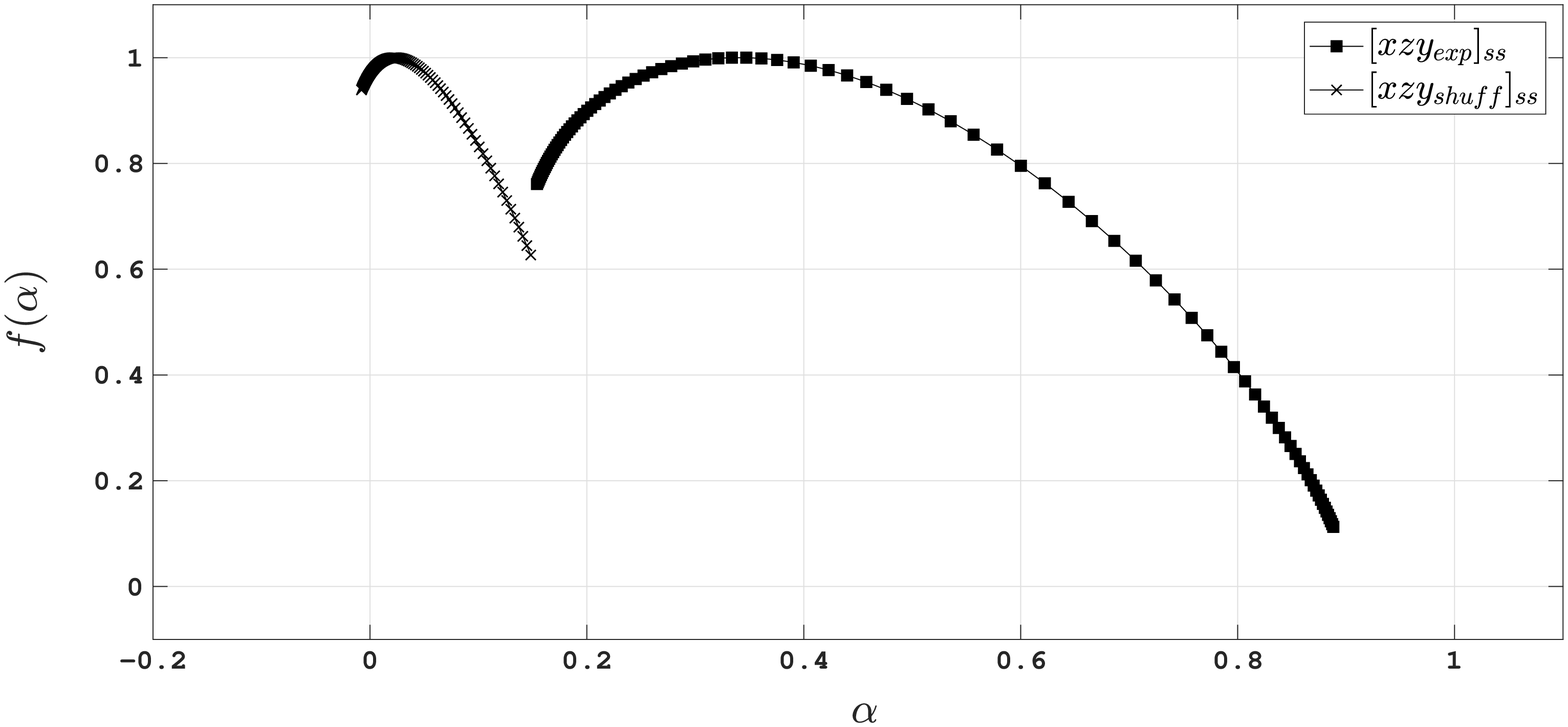}
\includegraphics[width=3.5in]{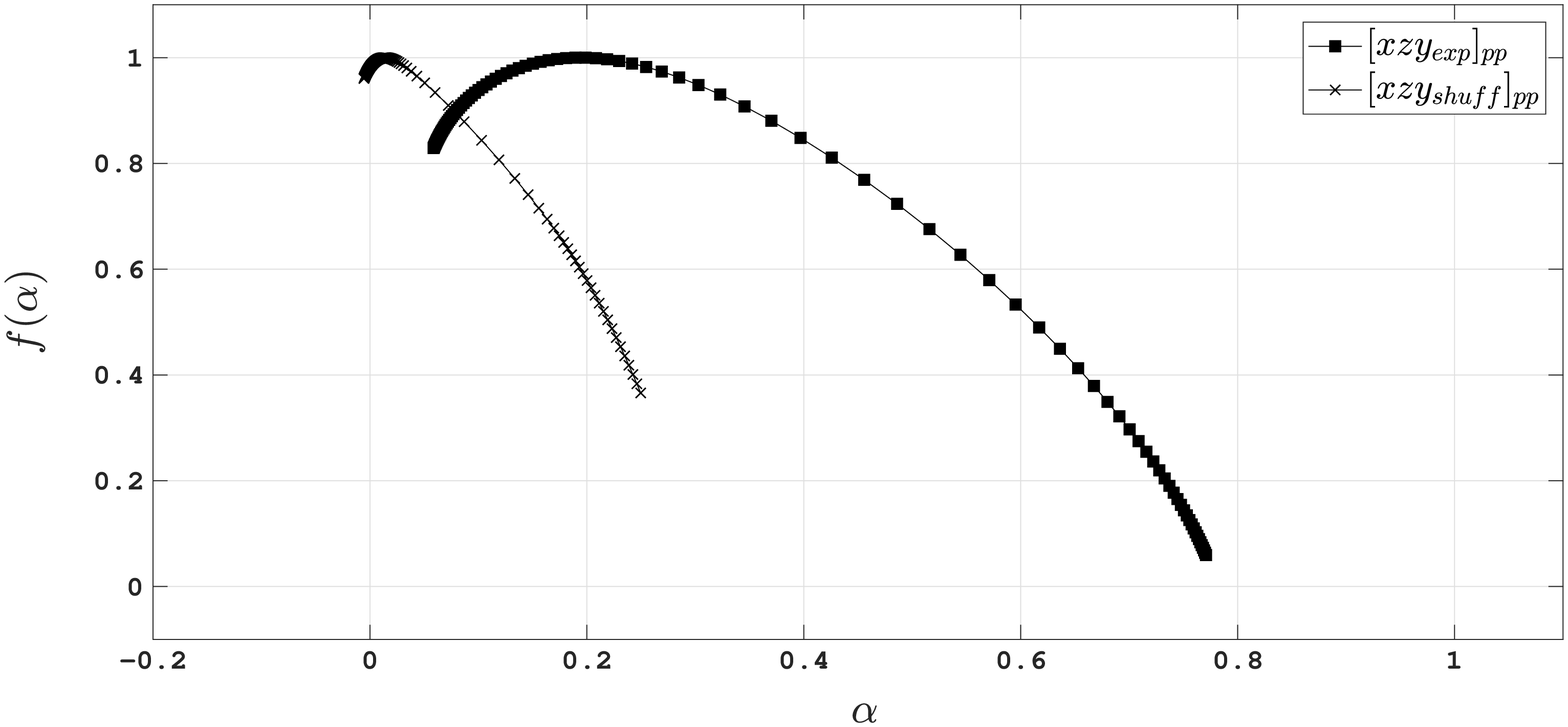}
}
\centerline{(a) \hspace*{6cm} (b)}
\caption{Comparison of the trend of different values of $f(\alpha)$ versus $\alpha$ between the experimental and the randomized version for the data surfaces (a) $[x,z,y]_{ss}$ (b) $[x,z,y]_{pp}$.}
\label{w_xzy}

\centerline{
\includegraphics[width=3.5in]{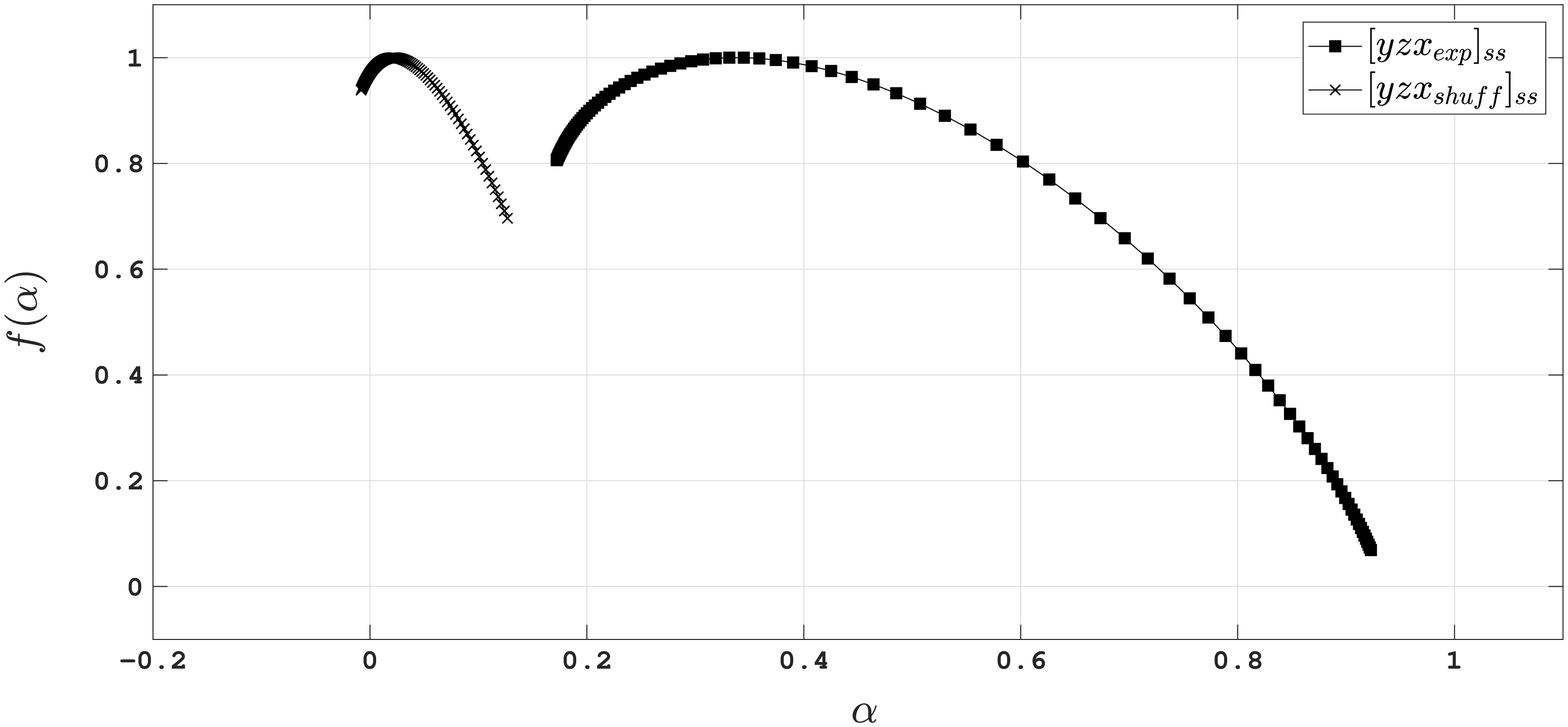}
\includegraphics[width=3.5in]{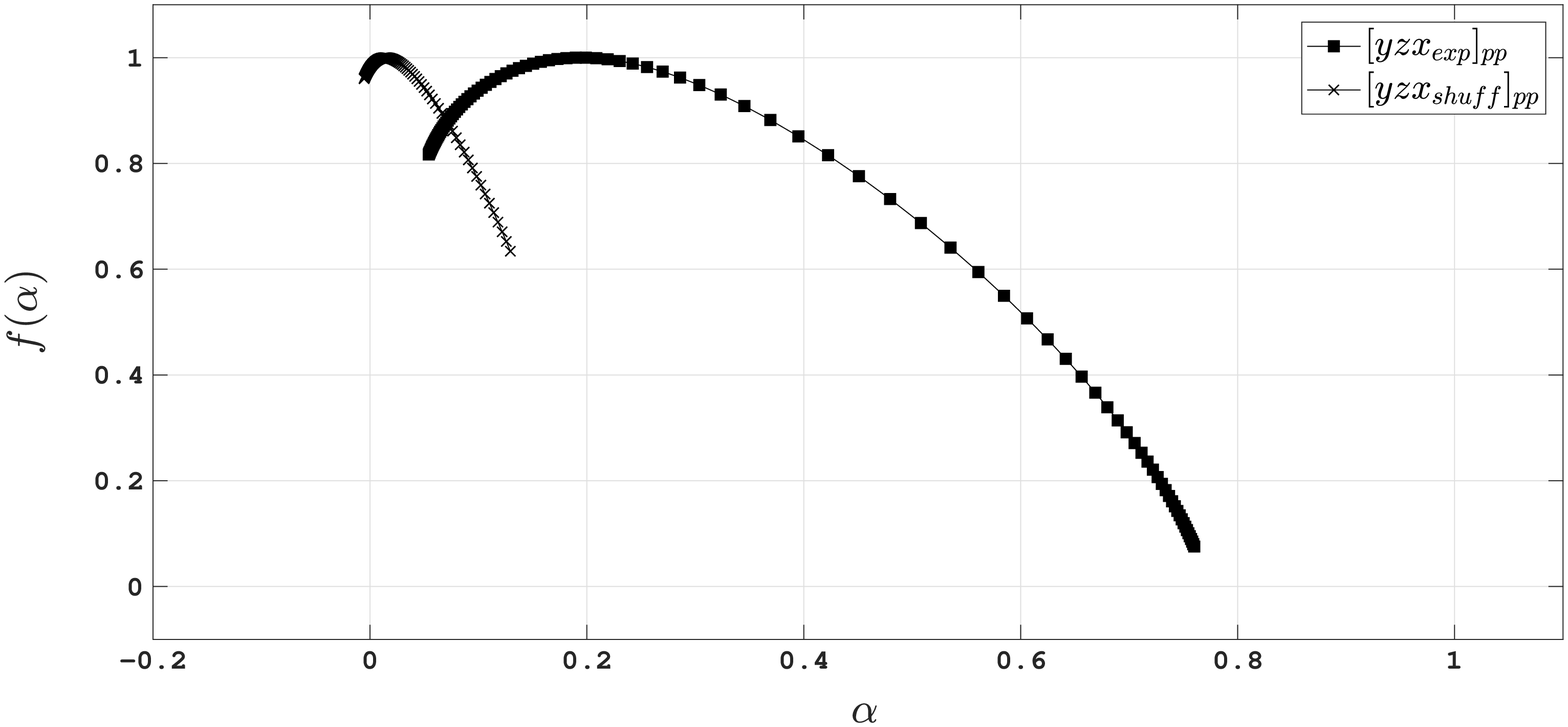}
}
\centerline{(a) \hspace*{6cm} (b)}
\caption{Comparison of the trend of different values of $f(\alpha)$ versus $\alpha$ between the experimental and the randomized version for the data surfaces (a) $[y,z,x]_{ss}$ (b) $[y,z,x]_{pp}$.}
\label{w_yzx}

\end{figure*}

In the Figures~\ref{w_xyz}-(a) and~\ref{w_xyz}-(b), the trends of different values of $f(\alpha)$ versus $\alpha$ extracted for the (2D) multifractal spectrum of the experimental data surfaces and their randomized versions, are shown for Super-Symmetric data surface $[x,y,z]_{ss}$ and non-Super-Symmetric data surface $[x,y,z]_{pp}$ respectively. The same trend for Super-Symmetric data surface $[x,z,y]_{ss}$ and non-Super-Symmetric data surface $[x,z,y]_{pp}$ are shown in the Figures~\ref{w_xzy}-(a) and~\ref{w_xzy}-(b). And for Super-Symmetric data surface $[y,z,x]_{ss}$ and non-Super-Symmetric data surface $[y,z,x]_{pp}$, the same the trends of different values of $f(\alpha)$ versus $\alpha$ are shown in Figures~\ref{w_yzx}-(a) and~\ref{w_yzx}-(b). The widths of (2D) multifractal spectrum for all the experimental data surfaces and their corresponding randomized versions are deduced as per step~\ref{seqn6} in Section~\ref{2dmfdfa} and listed in Table~\ref{mf_comp}.
\begin{table*}[h]
\refstepcounter{table}
\label{mf_comp}
\begin{center}
Table \arabic{table}
Widths of (2D) multifractal spectrum calculated for $6$ experimental data surfaces(matrices)[$3$ for supersymmetry data and $3$ for $pp$ collision data at $7$ TeV] and their randomized versions.\\
\begin{tabular}{@{}ccc@{}} 
\hline
\multicolumn{3}{c}{\textbf{Width of (2D) multifractal spectrum}}\\
\hline
&\textbf{$[x,y,z]_{ss}$}&\textbf{$[x,y,z]_{pp}$}\\
\cline{2-3}
\textbf{Original}&$0.760$&$0.640$\\
\hline
\textbf{Randomized}&$0.095$&$0.151$\\
\hline	
&\textbf{$[x,z,y]_{ss}$}&\textbf{$[x,z,y]_{pp}$}\\
\cline{2-3}
\textbf{Original}&$0.730$&$0.710$\\
\hline
\textbf{Randomized}&$0.156$&$0.255$\\
\hline
&\textbf{$[y,z,x]_{ss}$}&\textbf{$[y,z,x]_{pp}$}\\
\cline{2-3}
\textbf{Original}&$0.750$&$0.710$\\
\hline
\textbf{Randomized}&$0.135$&$0.135$\\
\hline
\end{tabular}
\end{center}
\end{table*} 
Comparison of the trend of different values of generalized Hurst exponent $h(q)$ versus $q$ for $q$ for $q=-5, \ldots 5$ between the Super-Symmetric and non-Super-Symmetric data surfaces constructed out from the $3$ perspectives - $[x,y,z]$, $[x,z,y]$, $[y,z,x]$ are shown in Figure~\ref{hq_q}-(a), (b) and (c) respectively. The values of generalized Hurst exponent $h(q)$ for $q=-5, \ldots 5$, are calculated as per the steps~\ref{rms_calc},~\ref{qfluc},~\ref{seqn4} in Section~\ref{2dmfdfa}.
\end{enumerate}

The observations and inferences from the Figures and the Table are listed below
\begin{itemize}
\item It must be noted that the shape of all the (2D) multifractal spectrum (figures~\ref{w_xyz},\ref{w_xzy} and \ref{w_yzx}), denoted by the trends of different values of $f(\alpha)$ versus $\alpha$, is non-symmetric. All the spectrum are truncated from the left which signifies that the trend of the generalized Hurst exponents-$h(q)$s calculated for positive values of $q(q>0)$-s for all the data surfaces, is almost consistent or uniform in nature. The almost uniform trend of generalized Hurst exponents for $q(q>0)$-s for the Super-Symmetric and non-Super-Symmetric data surfaces constructed out from the $3$ perspectives  - $[x,y,z]$, $[x,z,y]$, $[y,z,x]$, is evident from the Figure~\ref{hq_q}. This consistent trend of generalized Hurst exponents-$h(q)$s for $q(q>0)$-s indicate that the $q^{th}$ order fluctuation of fluctuation function $F_q(S(i))$ is not much affected by the extent of the local fluctuations for positive values of $q$. Also the (2D) multifractal spectrum have long right tails indicating the (2D) multifractal structures of the data surfaces are not affected by the higher magnitudes of the local fluctuations.

\begin{figure*}[h]
\centerline{
\includegraphics[width=3.5in]{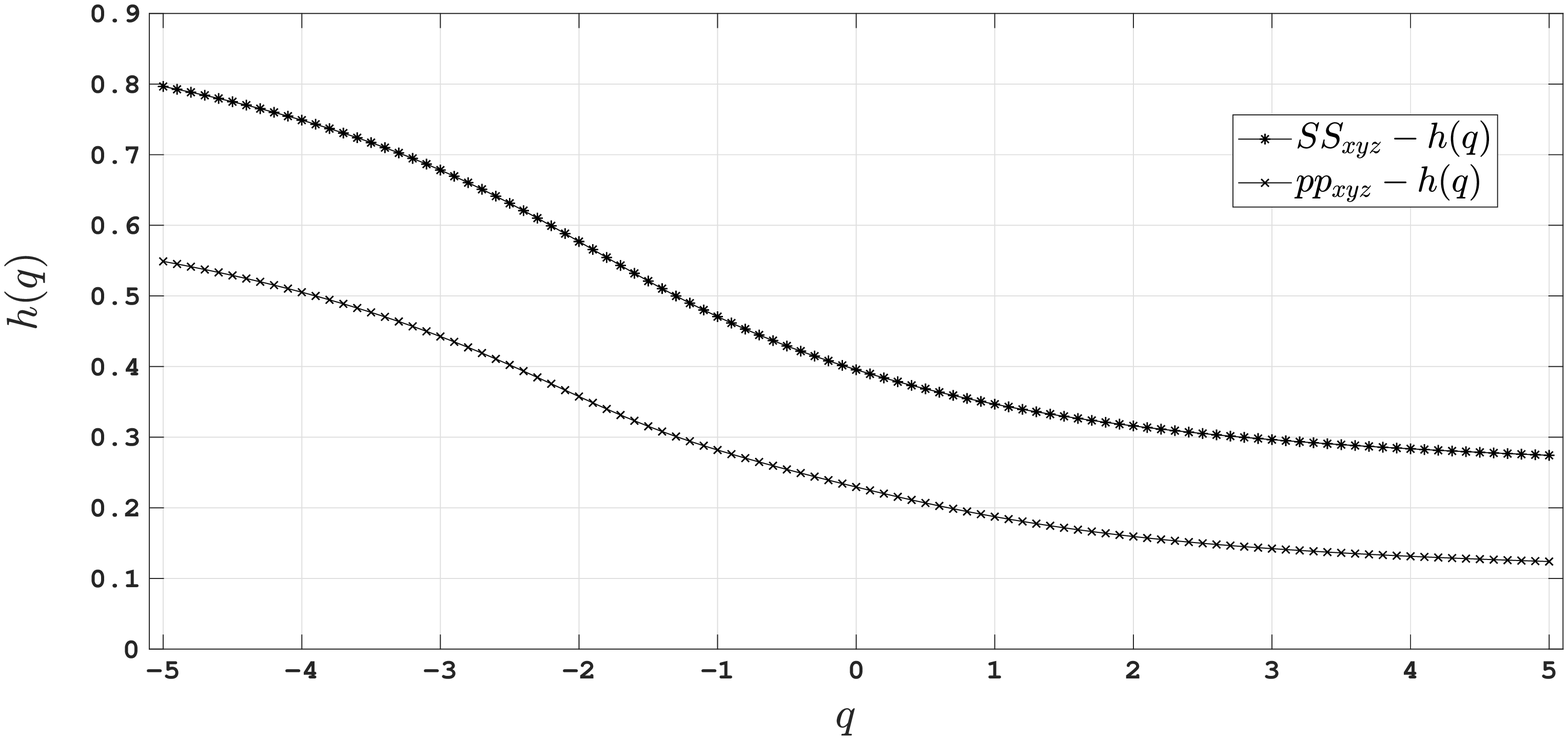}
\includegraphics[width=3.5in]{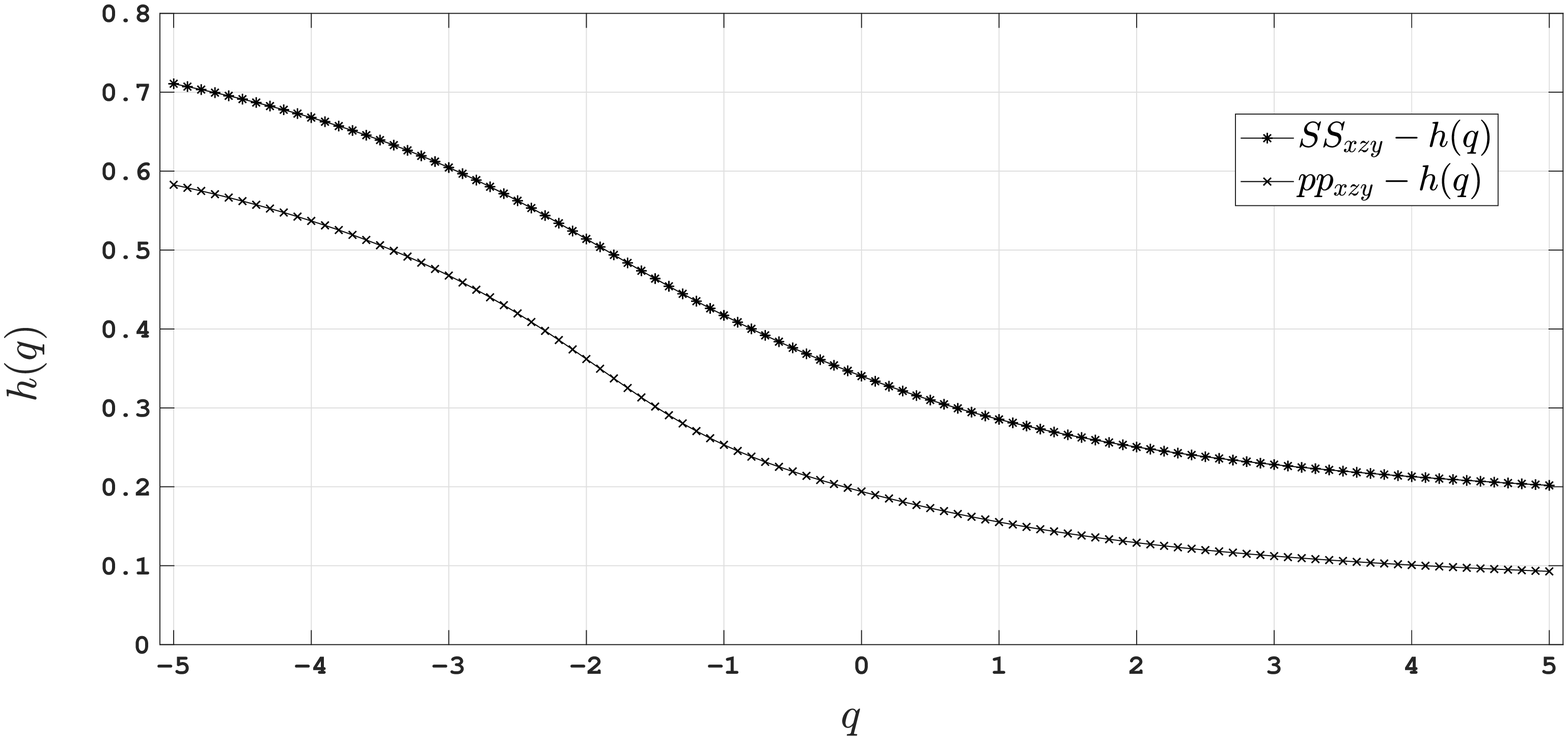}
}
\centerline{(a) \hspace*{6cm} (b)}
\centerline{
\includegraphics[width=3.5in]{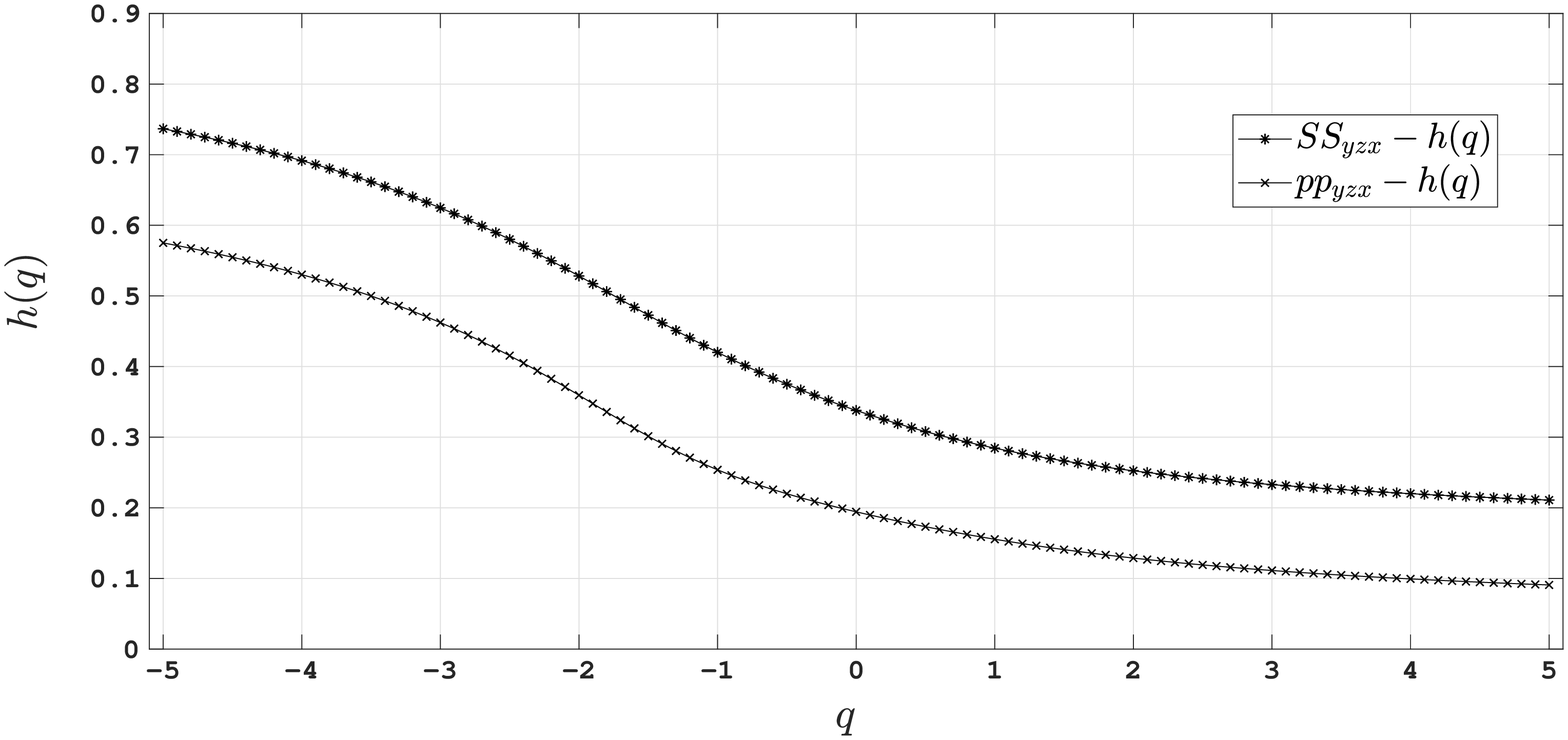}
}
\centerline{(c)}
\caption{Comparison of the trend of different values of generalized Hurst exponent $h(q)$ versus $q$ for $q$ for $q=-5, \ldots 5$ between the Super-Symmetric and non-Super-Symmetric data surfaces (a) $[x,y,z]_{ss}$ and $[x,y,z]_{pp}$ (b) $[x,z,y]_{ss}$ and $[x,z,y]_{pp}$ (c) $[y,z,x]_{ss}$ and $[y,z,x]_{pp}$}
\label{hq_q}
\end{figure*}

\item The long range correlation inherent in the experimental data, is eliminated by the process of randomization making the experimental data uncorrelated. This makes the width of (2D) multifractal spectrum signifying the trend of scaling pattern for different orders($q$-s), substantially less for the of randomized version of the data surfaces than the experimental ones. 

In this experiment the shape of the (2D) multifractal spectrum for randomized version is similar to that of experimental ones. This implies for both the experimental data surfaces and their randomized versions signifying the (2D) multifractal structures of the data surfaces and their randomized versions are not affected by the large values of the local fluctuations, as explained in the previous point. However, it's evident from the Table~\ref{mf_comp} that the widths of the (2D) multifractal spectrum of the experimental data is significantly more than that of their randomized versions. This establishes the statistical significance of the experimental result.

Hence, the widths of the (2D) multifractal spectrum computed for experimental data surfaces being noticeably different from their randomized ensembles, fundamentally establishes that the \textit{degree of scale-freeness} is manifestation of the inherent dynamics in the di-muon production process for both Super-Symmetric and non-Super-Symmetric data surfaces and not the characteristic of the dynamics of the randomized version and the values of the \textit{degree of scale-freeness} indicate various and unusual processes responsible for di-muon production 

\item It's evident from the Figure~\ref{hq_q} that there exists a consistently changing trend of $h(q)$ with $q$, though for $q<0$, $h(q)$ changes in a higher rate than for $q>0$. But this consistently changing behaviour of $h(q)$ with $q$ indicates the multifractal behaviour of the experimental data surfaces. Comparison of the trend of different values of generalized Hurst exponent $h(q)$ versus $q$ for $q$ for $q=-5, \ldots 5$ between the Super-Symmetric and non-Super-Symmetric data surfaces show that $h(q)$s for Super-Symmetric surfaces are around $50-100\%$ more than those calculated for the non-Super-Symmetric data surfaces for all $q$-s. 

\item Finally, if we compare the widths of the (2D) multifractal spectrum computed from the $3$ perspectives - $[x,y,z]$, $[x,z,y]$, $[y,z,x]$ for the range of $q=-5, \ldots 5$ for the Super-Symmetric data surfaces with those calculated for corresponding non-Super-Symmetric data surfaces in Table~\ref{mf_comp}, we find that the widths for Super-Symmetric data surfaces is around $10\%$(average) more than those of corresponding non-Super-Symmetric data surfaces.

\end{itemize}

\section{Conclusion}
\label{con}
For the past $60$ years Super-Symmetry has been been area of immense interest as it is the most encouraging theory in the are of high energy physics. 
The LHC has been enabling ATLAS and CMS collaborations to prove for it in various high energy interactions of $pp$ collisions at $7$, $8$, $13$ TeV and so on. Several hundreds of searches have been conducted to enrich our understanding of Nature.
For the past few decades, substantial amount of missing transverse momentum($p_T$) has been studied as the most potential observable for detecting the generation and decay of Super-Symmetric particles at different experiments at the colliders. At the LHC, the standard models of Super-Symmetry obeying the $R$-parity conservation predict signatures with jets originating from decaying pairs of squarks or gluinos, Super-Symmetric partners of quarks and gluons, and missing transverse momentum($p_T$) originating from undetected weakly communicating lightest Super-Symmetric particles or LSP. However, searches for such events have not been quite encouraging in displaying the signature of Super-Symmetry and also have not yielded strongest bounds as per the considered details of the Super-Symmetric models. The majority these bounds or limits obtained are considered as lower limit on the mass parameter. It has been concluded that Super-Symmetric partners must have masses higher than the limits deduced by the analysis.  However, searches for such events have not been quite encouraging in displaying the signature of Super-Symmetry and also have not yielded strongest bounds as per the considered details of the Super-Symmetric models. In summary, most of these decay processes share a distinct feature among themselves - the existence of many high energy jets originating from the expected high degree of difference between gluinos and the so-called daughter particles, and obviously, significant amount of missing transverse momentum($p_T$) yielding from the couple of high momentum LSP which escapes detection. 
At the LHC, there have been non-stop searches for Super-Symmetry for prompt as well as non-prompt, for R-parity conserving as well as R-parity violating generation and decays. From the results of these searches, it is inferred  that in these searches, LHC excludes the presence of gluinos below $2$ TeV, and existence of stops and gauginos below $1$ TeV. These checks or limits showed greater possibilities of the experiments in the collider. However, these signatures of Super-Symmetric particles are commonly derived under the assumption of a bit optimistic scenario where sparticles decay in to the final states under the study, with a $100\%$ branching fraction. The fact is that Super-Symmetry might have been in a disguised state at lower mass-scales as a results of difficult and challenging mass spectra and mixed modes of decays.

In the proposed investigation, we have extended (2D)(MF-DFA) proposed in~\cite{Yeh2012}, by selecting rectangular scale for doing Multifractal analysis for experimental data surfaces made up of the component-space(in the $X,Y,Z$ co-ordinates) taken out from the $4$-momenta of final state signatures of the produced di-muons for - Super-Symmetric data from the MultiJet primary $pp$ collisions dataset from RunB of 2010 at $7$ TeV and non-Super-Symmetric data from the primary dataset of $pp$ collisions at $7$ TeV from RunA of 2011 of the CMS collaboration, the details of which is written in Section~\ref{data}. The fundamental dynamics obtained from the signatures of final state particles of the di-lepton production process is reflected through the nonlinear and non-stationary parameters of symmetry based scaling analysis done using the novel method of extended (2D)(MF-DFA) method. The scaling exponent and other parameters differ substantially from non-Super-Symmetric to Super-Symmetric data which signifies that there is significant change in scaling behaviour from non-Super-Symmetric to Super-Symmetric data, as evident from the figures and table. This may be attributed to the occurrence of some unusual phenomena like Super-Symmetry. 

We have analyzed how the inherent pattern of scaling obtained from the signatures of final state particles has developed in the multi particle production process from non-Super-Symmetric to Super-Symmetric $pp$ collisions at $7$ TeV from the CMS collaboration and the inferences are summarised below.

\begin{enumerate}
\item The Table~\ref{hurst_comp} shows that the Hurst exponents denoted by $h(2)$ calculated from the slope of $log_2[F_q(S(i))]$ vs $log_2[S(i)]s$ for $q = 2$ obtained from the straight-line fitting for all the data surfaces (experimental and their randomised versions) from the $3$ perspectives - $[x,y,z]$, $[x,z,y]$, $[y,z,x]$, corresponding to both Super-Symmetric and non-Super-Symmetric $pp$ collision data at $7$ TeV, are all $<0.5$. This indicates that all the Super-Symmetric and non-Super-Symmetric data surfaces have inherent anti-persistent long-range correlations. Table~\ref{hurst_comp} shows that the values of the Hurst exponent calculated for the Super-Symmetric surfaces is significantly more (around $50\%$) than those of non-Super-Symmetric surfaces from the $3$ perspectives. This indicates that the anti-persistent long-range correlation and self-similarity inherent in the data surfaces changes significantly from non-Super-Symmetric surfaces to the Super-Symmetric ones and this is an unusual observation arising out of the dynamics of Super-Symmetric data.

Also, significantly different values of Hurst exponent between all the experimental data surfaces and their randomized versions establishes the statistical significance of the result of the experiment.

\item The linear trend of $log_2[F_q(S(i))]$ vs $log_2[S(i)]s$ for $q = -5,0,5$ calculated for the data surfaces from the $3$ perspectives - $[x,y,z]$, $[x,z,y]$, $[y,z,x]$, corresponding to both Super-Symmetric and non-Super-Symmetric $pp$ collision data at $7$ TeV, shown in the Figures~\ref{fqs_xyz},~\ref{fqs_xzy} and~\ref{fqs_yzx}, establish both fractal and multifractal nature of the data surfaces. This linear trend also confirms the self-similarity and long range power correlation of different orders of the experimental data surfaces.

\item It must be noted that all the (2D) multifractal spectrums generated from different values of $f(\alpha)$ versus $\alpha$ for all the experimental and their randomized version of both Super-Symmetric and non-Super-Symmetric data surfaces, are of similar non-symmetric shape. The (2D) multifractal spectrum have long right tails indicating the (2D) multifractal structures of the data surfaces are not affected by the higher magnitudes of the local fluctuations.
However, for Super-Symmetric and non-Super-Symmetric data, it's evident from the Table~\ref{mf_comp} and Figures~\ref{w_xyz}-(a) and~\ref{w_xyz}-(b) for $[x,y,z]$ perspective, Figures~\ref{w_xzy}-(a) and~\ref{w_xzy}-(b) for $[x,z,y]$ perspective and Figures~\ref{w_yzx}-(a) and~\ref{w_yzx}-(b) for $[y,z,x]$ perspective, that the widths of the (2D) multifractal spectrum of the experimental data is significantly different than that of their randomized versions. It indicates that long-range correlation and the \textit{degree of scale-freeness} inherent in the experimental data is completely different from the inherent dynamics of its randomised version. This establishes the statistical significance of the experimental result.

Also comparison of the widths of the (2D) multifractal spectrum computed for the Super-Symmetric data surfaces with those calculated for corresponding non-Super-Symmetric data surfaces, from the $3$ perspectives, listed in Table~\ref{mf_comp}, reveals that the structural difference between sections of small and large fluctuations of the Super-Symmetric data surfaces is more than that of corresponding non-Super-Symmetric data surfaces. This structural difference yields to around $10\%$(average) higher value of the widths of the (2D) multifractal spectrum computed for the Super-Symmetric data surfaces than the same computed for the non-Super-Symmetric data surfaces.

\item Interestingly, it must also be noted from the comparison of the trend of different values of generalized Hurst exponent $h(q)$ calculated for different $q$-orders for $q=-5, \ldots 5$ between the Super-Symmetric and non-Super-Symmetric data surfaces constructed out from the $3$ perspectives - $[x,y,z]$, $[x,z,y]$, $[y,z,x]$, that $h(q)$-s for Super-Symmetric data are consistently $50-100\%$ more than those of non-Super-Symmetric data. This difference is evident from the Figures~\ref{hq_q}-(a), (b) and (c). 

As we know that for multifractal data surfaces $q^{th}$ order $h(q)$ for $q>0$ reflects the scaling behaviour of the surface segments with large fluctuations and for values of $q<0$, $h(q)$ depicts scaling pattern of the surface segments with smaller fluctuations, hence we can confirm that the self-similarity and the scaling pattern reflected by the Hurst exponent or $h(2)$ and also generalized Hurst exponents or $h(q)$-s inherent in the non-Super-Symmetric data are substantially different than those of Super-Symmetric data. We may infer that this difference is yielded from the unusual dynamics of Super-Symmetric interactions for $pp$ collision at $7$ TeV from CMS collaboration, which is manifested in the component-space(in the $X,Y,Z$ co-ordinates) taken out from the $4$-momenta of final state signatures of the produced di-muons.

\end{enumerate}

We can conclude that the noticeable changes in scaling behaviour, as displayed in the figures and table, may be designated to the occurrence of various types of resonance-like states or other unusual phenomena. So, the proposed novel method of two-dimensional Multifractal analysis using \textit{rectangular scales} is a robust one and is rigorous enough to identify different types of resonance-like states or other unusual phenomena like Super-Symmetry which cannot be identified by the conventional method of analysing the spectrum of invariant mass/transverse momentum.
Moreover, in regards to the advantages of proposed method, we may emphasize that the proposed method can identify prospective Super-Symmetry which the conventional methods of analysing invariant mass spectrum/transverse momentum may miss, let alone the simplicity and novelty of the analysis methodology using the component-space(in the $X,Y,Z$ co-ordinates) taken out in totality, from the $4$-momenta of final state signatures of the di-muons produced in the $pp$ collision at $7$ TeV from CMS collaboration.

\section*{Acknowledgements} 
\label{ack}
We thank CERN, the CMS collaboration, and the CMS Data Preservation and Open Access(DPOA) team for making research-grade collider data available to the public. 

\end{document}